\documentclass[a4paper,fleqn]{cas-sc}

\usepackage[numbers]{natbib}
\usepackage{longtable}
\usepackage{multirow}
\usepackage{caption}
\usepackage{pdfpages}

\def\tsc#1{\csdef{#1}{\textsc{\lowercase{#1}}\xspace}}
\tsc{WGM}
\tsc{QE}
\tsc{EP}
\tsc{PMS}
\tsc{BEC}
\tsc{DE}

\begin{document}
\let\WriteBookmarks\relax
\def\floatpagepagefraction{1}
\def\textpagefraction{.001}

\shorttitle{Leveraging interaction data}

\shortauthors{et~al.}

\title [mode = title]{A Closer Look on Gender Stereotypes in Movie Recommender Systems and Their Implications with Privacy}                      




%

\author[1, 2]{Falguni Roy}[orcid=0000-0002-0427-5760]
\ead{falguniroy@hust.edu.cn, falguniroy.iit@nstu.edu.bd}
\cormark[1]

\affiliation[1]{
    organization={School of Computer Science and Technology, Huazhong University of Science and Technology}, 
    city={Wuhan 430074},
    country={China}
}
\affiliation[2]{
    organization={Institute of Information Technology, Noakhali Science and Technology University}, 
    city={Sonapur 3814},
    state={Noakhali},
    country={Bangladesh}
}

\author[3]{Yiduo Shen}
\ead{yiduo.shen@durham.ac.uk}
\affiliation[3]{
    organization={Department of Computer Science, Durham University}, 
    city={DH1 3LE},
    state={Durham},
    country={United Kingdom}
}

\author[1]{Na Zhao}
\ead{zhaona@hust.edu.cn}

\author[1]{Xiaofeng Ding}[orcid=0000-0001-5054-8515]
\ead{xfding@hust.edu.cn}

\author[4]{Md. Omar Faruk}
\ead{s224158311@deakin.edu.au}
\affiliation[4]{
    organization={Institute for Intelligent Systems Research and Innovation, Deakin University}, 
    city={75 Pigdons Rd, Waurn Ponds VIC 3216},
    country={Australia}
}

\begin{abstract}
The movie recommender system typically leverages user feedback to provide personalized recommendations that align with user preferences and increase business revenue. This study investigates the impact of gender stereotypes on such systems through a specific attack scenario. In this scenario, an attacker determines users' gender—a private attribute—by exploiting gender stereotypes about movie preferences and analyzing users' feedback data, which is either publicly available or observed within the system. The study consists of two phases. In the first phase, a user study involving 630 participants identified gender stereotypes associated with movie genres, which often influence viewing choices. In the second phase, four inference algorithms were applied to detect gender stereotypes by combining the findings from the first phase with users' feedback data. Results showed that these algorithms performed more effectively than relying solely on feedback data for gender inference. Additionally, we quantified the extent of gender stereotypes to evaluate their broader impact on digital computational science. The latter part of the study utilized two major movie recommender datasets: MovieLens 1M and Yahoo!Movie. Detailed experimental information is available on our GitHub repository: \textit{https://github.com/fr-iit/GSMRS.git}
\end{abstract}



\maketitle

\section{Introduction}

Nowadays, recommender systems are widely used to manage the overwhelming amount of information on the web. They help users by finding what they require while providing economic benefits to service providers. The recommender system even increases the decision-making efficiency of the users by delivering precise recommendations \citep{FENG2024103905, TIAN2023103815}. A recommender system (RS) typically consists of two major components, data and algorithm, which work together to produce recommendations. The data component provides the required formatted data for the algorithm. The algorithm then ranks items based on their relevance to users. Once users receive the ranked list of items, they interact with it by clicking, browsing, rating, or reviewing. These interactions, collectively called user feedback, become new data for the RS. This new data is combined with the existing data to improve future recommendations. The main workflow of the RS is cyclic.  

\begin{figure}[ht]
  \centering
  \includegraphics[width=0.75\textwidth]{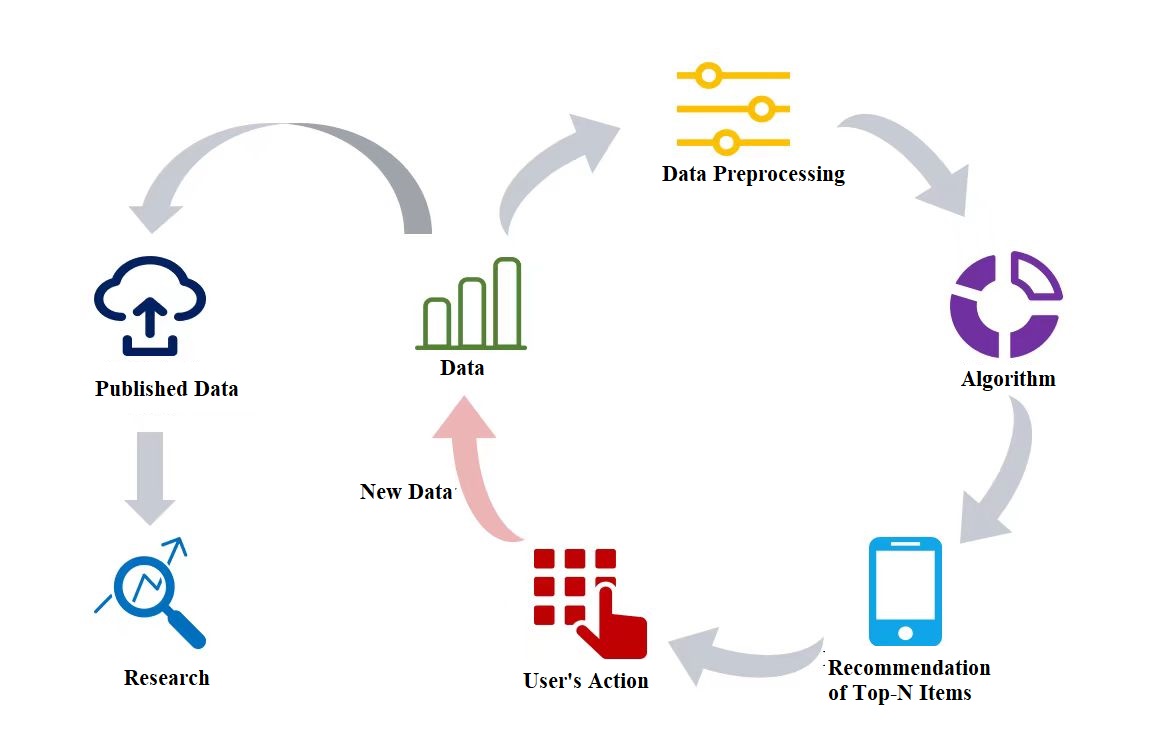}
  \caption{Simple Representation of Recommender System Process}
  \label{fig:RS}
\end{figure}

User feedback is categorized as implicit or explicit \citep{shi2023selection, wu2022adapting}. Implicit feedback includes the frequency of item views, clicks, and purchases \citep{BAI2024103525} and is usually assumed to represent positive user feedback. Conversely, explicit feedback includes ratings, reviews, and likes or dislikes provided directly by users. Explicit feedback reflects users' true feelings and is often considered more accurate than implicit feedback due to the active participation of users \citep{roy2023item}. Additionally, explicit feedback can capture users' positive and negative, high and low preferences. For example, in the context of the 1-5 rating range, the rating values $5$, $3$, and $1$ for an item $k$ indicate the user's high, neutral, and low preference on the item $k$, respectively. Similarly, for the like and dislike feedback method, a user's "like" or "dislike" for an item $k$ denotes their positive and negative preference toward the item. Among explicit feedback approaches, ratings are the most widely used form \citep{ricci2021recommender, reusens2017note}. Furthermore, the $data$ component of the RS has two main functions. First, it provides the resources needed to run the algorithm. Second, it publishes data for research purposes \citep{DENG2025DPDSA}. For example, websites such as MovieLens\footnote{https://movielens.org/}, Yahoo\footnote{https://www.yahoo.com/entertainment/movies/}, and Netflix\footnote{https://www.netflix.com/} are not only popular movie recommender systems but also publish their data\footnote{\label{ML}https://grouplens.org/datasets/movielens/},\footnote{\label{YM}https://webscope.sandbox.yahoo.com/},\footnote{https://www.kaggle.com/datasets/netflix-inc/netflix-prize-data} for research purposes. Even without publishing the dataset by the RS, the dataset can be created by crawling users' actions within the system \citep{ghasemi2021neural}, using transfer learning in cross-domain applications \citep{wang2021cross, cao2022contrastive} or bridging domain-specific information to target users \citep{khan2020predicting}. A simple demonstration of how a Recommender System works is presented in Figure \ref{fig:RS}.    

Usually, two types of information can be derived from user feedback: the user$-$item interaction matrix and the user's behavior history. The user$-$item interaction matrix contains implicit information about users, such as interaction patterns, demographic details, and personality traits \citep{base01_2021Gender, wang2021cross}. This implicit information creates a privacy vulnerability that adversaries can exploit to access sensitive user data. In the context of the movie recommender system (MRS), the user$-$item interaction matrix typically includes the user ID, item ID, and a numerical value representing the user's feedback. This feedback indicates the user's preference for the corresponding item. Implicit information can be extracted from the user$-$item interaction matrix by applying inference techniques with machine learning classifier algorithms \citep{base01_2021Gender, weinsberg2012blurme}.

Conversely, gender stereotypes (GSs) refer to the shared beliefs about specific genders regarding their consumption patterns, preferences, personality traits, proficiency, and behavior \citep{wuhr2017tears, russo2020gender}. These stereotypes not only influence expectations around gender-specific traits but also extend to the types of products, hobbies, and media content deemed suitable for each gender. In the context of movies, specific common beliefs exist regarding gender-oriented movie genre consumption and preferences \citep{wuhr2017tears}. These beliefs, denoted as GSs, are deeply rooted in societal norms and are reinforced through marketing strategies, media portrayal, and cultural expectations, collectively shaping film production and consumption patterns.

Gender stereotypes in movie preferences refer to the widely held assumptions about the types of movies men and women are expected to enjoy based on their gender. Societal norms often amplify these stereotypes, creating biases influencing how films are marketed, perceived, and consumed. These stereotypes can be conceptualized as interconnected networks linking gender and movie genres, where specific genres are strongly associated with either masculinity or femininity \citep{fabris2020gender}. Given the presence of gender stereotypes in movie preferences, this study aims to investigate how these stereotypes manifest in the context of movie recommendations and user behavior. Specifically, it seeks to answer the following questions:          

\begin{itemize}
    \item \textbf{RQ1:} What are the gender stereotypes in the case of movie preference? 
    \item \textbf{RQ2:} Does the gender stereotype exist in the movie recommender system?
    \begin{itemize}
        \item \textbf{RQ2.1:} If so, what is the magnitude of the presence of gender stereotypes in the movie recommender system? 
    \end{itemize}
\end{itemize}

In answering these questions, the paper makes the following four-fold contributions:

\begin{itemize}

    \item We define the gender stereotypes of movie preferences with the updated list of genres by conducting a user study with 630 samples\footnote{The authors can provide the original questionnaires \& anonymized data upon request\label{qus}}.
    
    \item We investigate gender stereotypes' existence in the movie recommender system with magnitudes of the GS's presence on the two popular publicly available datasets named MovieLens 1M (ML1M) and Yahoo!Movie by the state-of-the-art machine learning classifiers. We have chosen rating as the users' explicit behavior.   
    
    \item We also state the implications of gender stereotypes' presence in the context of privacy and bias issues of the recommender system.

    \item Our final contribution is sharing the experimental code used in our study. This allows other researchers to easily recreate our results and employ them as a standard for comparison.
    
 \end{itemize}

The rest of the sections are outlined as follows: Section \ref{sec:RW} demonstrates a list of existing literature that aligns with the present work. Section \ref{sec:method} explains the study's methodology. The experimental details with the results of the research questions are manifested in Section \ref{sec:EE}. In the last Section \ref{sec:cons}, we conclude the study by summarizing the findings and limitations by discussing the potential avenues for future research.

\section{Related Work}
\label{sec:RW}

\subsection{Research on Gender Stereotypes}

Researchers have investigated various domains to determine whether gender stereotypes (GSs) exist in the corresponding areas of study. They have identified a strong presence of GSs across diverse contexts, from surgeons and robots to daily life scenarios such as workplaces, homes, and digital behavior \citep{ashton2019stereotypes, eyssel2012s, wynn2018combating, fabris2020gender, ahn2022effect, wuhr2017tears}. For example, \citet{fabris2020gender}  examined the presence and impact of GSs on information retrieval systems and search engines. They explored gender bias encoded in word embeddings and its influence on the search results. Gender stereotypes were measured by applying implicit association tests on word embeddings. Word embeddings typically capture well-known stereotypes related to gender, and the genderedness score of a word is used to measure its association with masculinity or femininity. According to their analysis, models utilizing biased word embeddings for semantic analysis are highly susceptible to reinforcing GSs. However, neural models relying on original word representations can minimize this effect. 

Similarly, \citet{ahn2022effect} investigated the interactions between users and artificial intelligence (AI) agents in e-commerce, focusing on the effects of GSs on the perceived reliability of AI recommendations. Their findings indicated that individuals tend to associate competence with male AI agents and warmth with female AI agents. Additionally, in e-commerce, users express greater satisfaction with recommendations from female AI agents for hedonic products, while perceived male AI agents are more persuasive for valuable products. They also concluded that trust in AI recommendations varies based on the perceived gender of the AI agent. 

Furthermore, \citet{lange2021time} examined GSs in digital gaming and found that males prefer achievement-oriented, competitive, and aggressive games, while females favor music and dance games, quiz games, and educational games. According to their study, GSs regarding digital game genre preferences exist and can harm players by preventing them from pursuing their individual preferences. While some stereotypes in this context reflect fundamental differences between genders, others are inaccurate in size or scope. Moreover, \citet{wuhr2017tears} explored GSs in movie genre preferences, noting that males prefer aggressive, conflict-driven genres, whereas females prefer movies emphasizing social relationships. Their study confirmed the existence of GSs in digital media, particularly in movie preferences, and claimed that these stereotypes are accurate in direction.

\subsection{Research on Private Data Inference}
\label{sec:GI}

Many researchers have studied and identified the correlation between individuals' public and private data in the digital domain. Public data typically refers to browsing or purchase history, ratings, reviews, and like or dislike information on the web. Conversely, private data refers to information that individuals do not willingly share or disclose online, such as income, age, gender, or political opinions. Most people remain unaware of the link between their public and private information \citep{salamatian2015managing}. A study by \citet{kosinski2013private} demonstrated that using Facebook likes, a wide range of sensitive information about a user, such as age, gender, personality traits, intelligence, happiness, and parental separation, can be predicted with 85-95\% accuracy. Similarly, \citet{bi2013inferring} investigated the acquisition of demographic traits, such as age, gender, and even political and religious views, from search queries, achieving up to 80\% accuracy. Additionally, \citet{feng2014tags} showed that users' private data can be identified from their implicit viewing history on online video systems.

However, private data may vary among individuals. For instance, some people share their living location on social networks, while others consider it private. \citet{jia2017attriinfer} illustrated that it is possible to infer a user's private attributes, such as location and interests, using data from a fraction of other users who share this information on social networks by applying regression algorithms. Furthermore, the context of private data may differ across recommender systems. For example, a user might disclose personality information in a tourism recommender system to receive precise recommendations but may choose not to reveal the same data on an e-commerce website.

Nevertheless, it is possible to gain the user's private data by applying transfer learning knowledge in the cross-domain. \citet{wang2021cross} proposed a framework to solve the personality trait classification problem in a cross-domain setting using predictive text embedding for transfer learning. The authors deduced users' personality traits in the target domain by leveraging transfer learning from the source domain. They further suggested utilizing this information to enhance the performance of target recommender systems by improving prediction accuracy. 

In some cases, users' private data can be inferred using only the rating values they provide in a system, which indicate their degree of preference and satisfaction with consumed items. \citet{weinsberg2012blurme} pioneered the study of private data inference, focusing on gender as private data, and achieved 80\% inference accuracy. They also proposed a solution to address privacy threats identified in their study. Later, \citet{feng2015can} extended this approach to infer users' gender and age and developed a module between users and recommender systems to preserve privacy. Similarly, \citet{salamatian2015managing} applied inference algorithms to detect users' political views based on their history of watching political or TV shows and their ratings. Following this approach, \citet{base01_2021Gender} focused on inferring users' gender from rating data and proposed a solution to preserve the privacy of the entire user-item interaction matrix. 

Existing studies on recommender systems mainly focus on enhancing recommendation performance in terms of accuracy by integrating users' public and private data or addressing privacy concerns through proposed solutions. However, limited attention has been given to understanding the root causes of privacy vulnerabilities while users' public and private data contain strong correlations. As a consequence, this study aims to fill this gap by investigating whether GSs embedded in movie preferences can be leveraged to infer users' private attributes, such as gender, from their interaction data within Movie Recommender Systems (MRS). Unlike previous works, which often propose solutions for privacy preservation of the system, this research focuses solely on identifying the existence and extent of the problem. To achieve this, we first identified potential GSs associated with various movie genres through a user study. Then, we verified their presence as the preference of the items consumed by users, using explicit ratings as feedback to compile users' interaction data. Afterward, we quantified the prevalence of GSs within the system to determine their impact on user privacy. Our findings indicate that the existence of GSs in MRS data increases the likelihood of adversaries inferring users' private data by correlating stereotypes with their preferences. While this study does not propose solutions, it provides a critical foundation for future research to develop privacy-preserving mechanisms by considering the risks posed by GSs in recommender systems.

\section{Study Design \& Methodology}
\label{sec:method}

This section contains three subsections. In the first subsection, we set gender-wise hypotheses to measure the movie genre preferences as GSs. In the consecutive subsection, we apply the output of the previous subsection in the different inference algorithms to find the presence of GSs in the MRS, and the magnitude of the presence of GSs is measured in the last subsection. 


\subsection{Gender Stereotypes Detection for Movie Preference}
\label{sec:GSD}

Watching movies is one of the most popular forms of entertainment that people often select for amusement. Usually, movies contain strong characteristics that can impact audiences by providing a social message and introducing moral values and cultural differences. Additionally, movies can be a tool for quickly learning and understanding different languages and may even play a role in psychotherapy \citep{correia2018cinema}.  The existing literature shows that movie selection primarily depends on people's demographic (gender, age, and personality) differences, genre preferences, consumption frequency, and cognitive and affect elements, such as psychographics \citep{palomba2020consumer}. In addition, demographic information is an important parameter for advertising and marketing campaigns, as it aids in understanding genre preferences and consumption frequency \citep{kim2018demographic}. Typically, for watching a movie, females are supposed to watch romance, relationships, and emotional themes movies, which focus on interpersonal relationships, love stories, and personal growth. These preferences reflect an inclination toward movies focusing on emotional connections between characters. Conversely, males are perceived to watch "guy flicks" movies. The "guy flicks" movies are generally focused on fast-paced action sequences and fear enjoyment, such as car chases, explosions, physical feats, and shocking visuals that raise fear feelings \citep{wuhr2017tears, palomba2020consumer, infortuna2021inner, martin2019you}. 

Genre preferences are most strongly influenced by gender, often aligning with traditional GSs \citep{grodal2017film}. The approximate origins of GSs in movie preferences can be attributed to media content, biological factors, and social theory \citep{wuhr2017tears}. For example, by considering biological accounts, testosterone, a hormone predominantly found in males, is associated with behaviors characterized by dominance \citep{procyshyn2020experimental}. This correlation could explain why males are more interested than females in movies centered around competition and action-adventure themes. On the other hand, the oxytocin hormone, primarily associated with females, is believed to foster pair bonding and generate feelings of love in humans \citep{procyshyn2020experimental}. From a neurobiological standpoint, this distinction might underlie females' heightened inclination toward relationships and love story category' movies. Accordingly, we set the following hypotheses to define the GSs for movie genre preference to determine the answer to the $RQ1$. 

\begin{itemize}
    \item \textbf{H1: } Females prefer movies centered on social relationships more than males, especially those affiliated with romance, drama, and family genres.
    \item \textbf{H2: } Males prefer movies centered on aggressive conflicts \& dominant themes more than females, particularly in genres such as horror, action, adventure, fantasy, crime, and war.
\end{itemize}

\subsection{Gender Inference Algorithm}
\label{sec:GIA}

Generally, when selecting a movie to watch, users first choose a genre and then select a movie within that genre \citep{palomba2020consumer, steck2018calibrated}. Additionally, demographic factors, such as gender, significantly influence the selection of movie genres \citep{palomba2020consumer, bolock2020towards}. Based on these observations, we propose the following threat model:

\textbf{Adversary: } 
A party interested in users' private attributes could include advertisers, banking institutions, insurance companies, or cybercriminals \citep{jia2017attriinfer}.

\textbf{Threat Model: } 
\label{threatmodel} 
Assume that the adversary has access to a gender classifier and the necessary data, along with gender stereotypes, to train the classifier. The adversary's objective is to infer users' gender attributes. To achieve this, the adversary constructs a user-item interaction matrix using the available data. By leveraging the interaction matrix and gender stereotypes, the adversary can potentially deduce the target users' gender information.

\textbf{DEFINITION:  (\textit{Gender Inference Problem})}. Suppose an adversary has 1) binary gender attributes, 2) user-item interaction data, 3) names or genres of items, 4) knowledge about gender stereotypes of movie preferences, 5) a set of target users, and 6) a classifier. The gender inference aims to deduce whether the gender of each target user is male or female. The mathematical formulation of the gender inference problem is shown in Equation \ref{eq:GIP}:


\begin{equation}
\label{eq:GIP}
\begin{matrix}
[R_{UI} \; | \;  GS_{M,F}] \xrightarrow[u  \in U]{classifier} u_G \;\;\;\;\;\;\;
where, \;\;\; G \; \in \; M\;||\;F \\\\
GS_{M,F} = d_{M} \cup d_{F}
\end{matrix}    
\end{equation}

Here, $R_{UI}$ denotes user-item interaction data. $GS_{M,F}$ indicates GSs of movie preferences for the users and $u_G$ expresses the gender of the target user, which can be either male ($M$) or female ($F$). $U$ implies the set of target users. Furthermore, to detect the existence of GSs in the MRS by taking into account the threat model (Equation \ref{eq:GIP}), we first measure the degree of male and female genre preference ($d_{M}$ and $d_{F}$) as per GSs in all rated movies for a user by using the Equation \ref{eq:FMPrefer}. Actually, $d_{M}$ and $d_{F}$ indicate the extent to which a user consumes items from genres associated with the corresponding GSs. Afterward, the degree of male and female genre preferences and the user-item interaction matrix are used as feature vectors to train the gender classifier.

\begin{equation}
\label{eq:FMPrefer}
\begin{matrix}
d_{M} = \sum_{i = 1}^{I} g_i \cap g_{GS_M} \\ \\
d_{F} = \sum_{i = 1}^{I} g_i \cap g_{GS_F} 
\end{matrix}
\end{equation}

Here, $I$ denotes the consumed items set, and $i$ indicates each item, where $i \in I$, respectively. $g_i$ defines the list of genres belonging to item $i$, whereas $g_{GS_M}$ and $g_{GS_F}$ denote the list of genres associated with GS of male and female.  

\subsection{GSs Presence Measurement in RS}
\label{sec:GPM}

The value of the area under the curve (AUC) ranges between 0 and 1. If the AUC value for the gender inference algorithm is greater than 0.50, it indicates that the algorithm can distinguish between the classes. This also suggests that the data implicitly contains GSs related to movie preferences. To assess the degree of GS presence, we compare the proportions of male and female genre preferences with the actual gender data of the users using Equation \ref{eq:GS_Count}.

\begin{equation}
\label{eq:GS_Count}
\begin{matrix}
u_{GS} = (1- \frac{K}{U})*100 \;\;\;\; where \\ \\
K = \sum_{k=1}^{U} g((k_G = F \land d_{M_k}>d_{F_k}) \lor (k_G= M \land d_{M_k}<d_{F_k})) \\
\end{matrix}
\end{equation}

Here, $U$ represents the group of users, and $k$ denotes each user, where $k \in U$. The variable $k_G$ indicates the user's gender, which in this study is treated as a binary attribute denoting either male ($M$) or female ($F$). The term $u_{GS}$ represents the percentage of users whose preferences align with gender stereotypes. Additionally, $K$ refers to the count of users whose preferences deviate from gender-based stereotypes, as defined by the indicator function $g(.)$ expressed as follows:

\begin{itemize}
    \item For female users ($k_G = F$), it checks if their male genre preference $d_{M_k}$ is greater than their female genre preference $d_{F_k}$.
    \item For male users ($k_G = M$), it checks if their female genre preference $d_{F_k}$ is greater than their male genre preference $d_{M_k}$. 
\end{itemize}
 
\section{Experiments \& Evaluation}
\label{sec:EE}

\subsection{Methods and results for RQ1}
\label{sec:RQ1}

\begin{itemize}
    \item[] \textbf{RQ1: } \textit{What are the gender stereotypes in the case of movie preference?}
\end{itemize}

We conducted a user study to address our first research question (RQ1). The list of movie genres is updated frequently, so the current genre list does not align with prior gender-genre association studies \citep{wuhr2017tears}. To ensure relevance, we designed a user study based on the list of popular movie genres available on IMDb\footnote{\label{list}https://www.imdb.com/feature/genre}, an internationally recognized platform for movie recommendations that includes movie data from various nations. The list\footref{list} contains 23 genres. But some of the genres, such as music and musical, are almost identical with minor differences that we merged and denoted as 'musical'. Additionally, we excluded the 'short' genre from our list, as the duration of a movie typically does not significantly impact its entertainment value. Consequently, we considered a total of 21 genres for our study. The final list of genres is shown in Table \ref{tab:movieGen}. 

\begin{table}[ht]
\centering
\caption{List of Movie Genres}
\label{tab:movieGen}
\begin{tabular}{ccccccc}
\hline
\multicolumn{7}{c}{\textbf{Movie Genres}} \\ \hline
Action & Comedy & Family & Sports & Mystery & Sci-Fi & Horror \\
Adventure & Crime & Fantasy & War & Romance & Thriller & Musical (Music, Musical) \\
Animation & Documentary & Film-Noir & Western & Biography & Drama & History \\
\hline
\end{tabular}
\end{table}

We developed a questionnaire consisting of 32 items divided into three segments\footref{qus}. The first segment, containing items 1–4, collected participants' demographic data, such as age, gender, country, and occupation. In the next segment, items 5–25 focused on each of the 21 genres, asking participants to indicate their individual preferences on a 5-point rating scale. Precisely, each item was phrased as: “\textit{How much do you like following genre's (action) movies and affect your preference for movie selection on a scale from 1 (not at all) to 5 (extremely)?}". Following this, the final segment began with item 26, which measured an individual's movie consumption frequency. Items 27 and 28 assessed an individual's willingness to embrace a serendipitous nature. Subsequently, items 29 and 30 were included to gauge whether the participants were engaged or unengaged in the survey. Item 31 aimed to determine the most suitable way for participants to express their feelings about a movie. Lastly, item 32 evaluated the relevance of the participants' genre preferences. 

We ran the user study from November 2023 to mid-March 2024 (4.5 months). The study had two versions: one was the English language, and another was the Chinese language. Data were also collected through two sources: Google Forms for the English version and the mikecrm\footnote{www.mikecrm.com} website for the Chinese version. Participants were recruited voluntarily by sharing the questionnaire through social media such as WeChat, Facebook, WhatsApp and emails. To participate in the study, individuals needed to be familiar with the IMDb website and have previously watched at least 20 movies. This requirement assessed real male and female perceptions of gender-specific movie stereotypes. Additionally, participants had to list their top 10 favorite movies to compare the genres of these movies with the study's genre preferences, helping to identify any fake responses. All data collection was conducted anonymously. Only age, gender, occupation, and country were recorded by the participants, and no other personal information (such as email or name) was collected to prevent further tracking of the participants. Before completing the questionnaire, all the relevant information about the study was provided to the participants to give them a clear view of their provided data usage. In total, 677 people participated in the survey, but 47 unengaged responses were detected and removed after screening. At the end of the data screening process, 630 samples were finally chosen as data for successive experiments. Table \ref{tab:demodata} demonstrates the information about the study data.


\begin{table}[!ht]
    \centering
    \caption{Participants' Gender, Age \& Occupation Information of the Study}
    \label{tab:demodata}
    \begin{tabular}{llcc}
    \hline
    \multicolumn{2}{l}{\textbf{General Information}} & \textbf{Frequency} & \textbf{Percent} \\ 

    \hline
    Gender & Female & 229 & 36.3 \\ 
         ~ & Male & 401 & 63.7 \\ 
    \hline
    Age & 15-20 & 20 & 3.2 \\ 
        ~ & 21-25 & 255 & 40.5 \\ 
        ~ & 26-30 & 141 & 22.4  \\ 
        ~ & 31-35 & 149 & 23.7  \\ 
        ~ & 36-40 & 51 & 8.1  \\ 
        ~ & 41-45 & 11 & 1.7  \\ 
        ~ & 46-50 & 03 & 0.5  \\  
    \hline
    Occupation & Accounts Officer & 03 & 0.5  \\ 
        ~ & Architect & 02 & 0.3  \\ 
        ~ & Artist & 01 & 0.2  \\ 
        ~ & Bank Officer & 16 & 2.5  \\ 
        ~ & Business & 04 & 0.6  \\ 
        ~ & CEO & 01 & 0.2  \\ 
        ~ & Cyber Security Specialist & 01 & 0.2  \\ 
        ~ & Doctor & 04 & 0.6  \\ 
        ~ & Entrepreneur & 02 & 0.3  \\ 
        ~ & Financial Advisor & 01 & 0.2  \\ 
        ~ & Government Employee & 07 & 1.1  \\ 
        ~ & Manager & 04 & 0.6  \\ 
        ~ & Project Manager & 05 & 0.8  \\ 
        ~ & Research Assistant & 36 & 5.7  \\ 
        ~ & Senior Executive & 03 & 0.5  \\ 
        ~ & Service Holder & 14 & 2.2  \\ 
        ~ & Software Engineer & 50 & 7.9  \\ 
        ~ & SQA & 06 & 1.0  \\ 
        ~ & Student & 293 & 46.5  \\ 
        ~ & System Engineer & 08 & 1.3  \\
        ~ & Teacher & 165 & 26.2  \\
        ~ & Unemployed & 04 & 0.6  \\
    \hline
    \end{tabular}
\end{table}

\subsubsection{\textbf{Dependent Variable}} For this study, we consider gender as the dependent variable, which has binary attributes: male and female.

\subsubsection{\textbf{Explanatory Variables}} The genres are considered explanatory variables in this study. We collect the genre's preference value in a 5-point rating scale. Later, we convert it into three classes of category where the rating value "\textit{Not at all}" is associated with "\textbf{No}" class, and the rating value "\textit{Slightly}" or \textit{Moderately} belongs to "\textbf{MinPefer}" class. The rating value "\textit{Very}" or \textit{Extremely} is affiliated with the "\textbf{MaxPefer}" class. As a result, we have 21 explanatory variables, each containing three category values: MaxPrefer, MinPrefer, and No.

\subsubsection{\textbf{Genres That Influence A Viewer's Movie Preference: Distinguish by Gender}}

Since the dependent variable in our study is binary, we employed binary logistic regression as a multivariate analysis technique to identify the genres that influence viewers' movie preferences. The analysis was conducted using SPSS, a widely recognized statistical tool, and we utilized version 26 for this purpose. We repeated the same analysis twice to evaluate the genre influence separately for male and female viewers. Table \ref{tab:maleLG} presents the analysis result for the male viewers. The findings indicate that the \textbf{action, adventure, comedy, crime, horror,} and \textbf{war} genres have a significant positive impact (p < 0.05) on male viewers' movie preferences, while the \textbf{animation, family, drama,} and \textbf{romance} genres show a significant negative influence (p < 0.05). According to Table \ref{tab:maleLG}, males predominantly favor the \textbf{action} and \textbf{adventure} genres, enjoying them 5.42 times (odds = 5.417, p < 0.05) \& 3.85 times (odds =  3.849, p < 0.05) more than females, respectively. 

Additionally, males demonstrate a strong preference for \textbf{comedy, crime}, and \textbf{war} genres, preferring them 3.66 times (odds = 3.665, p < 0.05), 3.65 times (odds = 3.65, p < 0.05), and 4.87 times (odds = 4.867, p < 0.05) more than females when selecting movies to watch. Furthermore, our analysis reveals that males show an exceptionally high preference for the \textbf{horror} genre, with an odds ratio of 1.96 for maximum preference (p < 0.05) and 2.45 for minimum preference (p < 0.05) compared to females.

On the other side, the movie which belongs to \textbf{animation, drama} and \textbf{romance} genres are 64\% (odds = 0.359, p< 0.05),  59.5\% (odds = 0.405, p< 0.05) \& 60\% (odds =  0.396, p< 0.05) less preferred by males accordingly than female at the time of movie selection. Also, males do not like to watch \textbf{family} genre movies as this genre negatively impacts their movie selection motive for both the max (odds = 0.15, p< 0.05) \& min (odds =  0.19, p< 0.05) preference categories. Furthermore, the Chi-square statistic and p-value from the goodness of fit test (Chi-square: 532.125, p-value = 0.942) suggest that the binary logistic regression model fits the data well and yields more suitable outcomes.

{
\small
\begin{longtable}{lcccccc}
\caption{\newline Multivariate Binary Logistic Regression Defining Influential Movie Genres of Males for Movie Preferences} \label{tab:maleLG} \\
\toprule
\multicolumn{1}{c}{\multirow{2}{*}{\textbf{Variables}}} & \multirow{2}{*}{\textbf{\begin{tabular}[c]{@{}c@{}}Logistic \\ Coefficient\end{tabular}}} & \multirow{2}{*}{\textbf{\begin{tabular}[c]{@{}c@{}}Standard \\ Error\end{tabular}}} & \multirow{2}{*}{\textbf{\begin{tabular}[c]{@{}c@{}}Odds \\ Ratio\end{tabular}}} & \multicolumn{2}{c}{\textbf{95\% CI}} & \multirow{2}{*}{\textbf{p-value$^b$}} \\ \cline{5-6}
\multicolumn{1}{c}{} &  &  &  & \textit{\textbf{\begin{tabular}[c]{@{}c@{}}Lower Bound\end{tabular}}} & \textit{\textbf{\begin{tabular}[c]{@{}c@{}}Upper Bound\end{tabular}}} &  \\ 
\midrule
\endfirsthead
\caption[]{\newline Multivariate Binary Logistic Regression Defining Influential Movie Genres of Males for Movie Preferences (continued)} \\
\toprule
\multicolumn{1}{c}{\multirow{2}{*}{\textbf{Variables}}} & \multirow{2}{*}{\textbf{\begin{tabular}[c]{@{}c@{}}Logistic \\ Coefficient\end{tabular}}} & \multirow{2}{*}{\textbf{\begin{tabular}[c]{@{}c@{}}Standard \\ Error\end{tabular}}} & \multirow{2}{*}{\textbf{\begin{tabular}[c]{@{}c@{}}Odds \\ Ratio\end{tabular}}} & \multicolumn{2}{c}{\textbf{95\% CI}} & \multirow{2}{*}{\textbf{p-value$^b$}} \\ \cline{5-6}
\multicolumn{1}{c}{} &  &  &  & \textit{\textbf{\begin{tabular}[c]{@{}c@{}}Lower Bound\end{tabular}}} & \textit{\textbf{\begin{tabular}[c]{@{}c@{}}Upper Bound\end{tabular}}} &  \\ 
\midrule
\endhead
\bottomrule
\multicolumn{7}{r}{{Continued on next page}} \\ 
\endfoot
\endlastfoot

\textbf{{[}Action=MaxPrefer{]}} & \textbf{1.69} & \textbf{0.49} & \textbf{5.417} & \textbf{2.073} & \textbf{14.157} & \textbf{*0.001} \\ 
{[}Action=MinPefer{]} & 0.869 & 0.529 & 2.385 & 0.846 & 6.722 & 0.1 \\ 
{[}Action=No{]} & \multicolumn{6}{l}{\textit{***$Ref^a$}} \\ \hline

\textbf{{[}Adventure=MaxPrefer{]}} & \textbf{1.348} & \textbf{0.541} & \textbf{3.849} & \textbf{1.333} & \textbf{11.111} & \textbf{*0.013} \\ 
{[}Adventure=MinPefer{]} & 0.555 & 1.001 & 1.741 & 0.245 & 12.392 & 0.58 \\ \hline

\textbf{{[}Animation=MaxPrefer{]}} & \textbf{-1.024} & \textbf{0.388} & \textbf{0.359} & \textbf{0.168} & \textbf{0.769} & \textbf{*0.008} \\ 
{[}Animation=MinPefer{]} & -0.687 & 0.393 & 0.503 & 0.233 & 1.087 & 0.081 \\ \hline

{[}Biography=MaxPrefer{]} & 0.506 & 0.885 & 1.659 & 0.293 & 9.405 & 0.567 \\ 
{[}Biography=MinPrefer{]} & 1.036 & 0.861 & 2.818 & 0.521 & 15.238 & 0.229 \\ \hline

\textbf{{[}Family=MaxPrefer{]}} & \textbf{-1.899} & \textbf{0.821} & \textbf{0.15} & \textbf{0.03} & \textbf{0.749} & \textbf{*0.021} \\ 
\textbf{{[}Family=MinPefer{]}} & \textbf{-1.66} & \textbf{0.808} & \textbf{0.19} & \textbf{0.039} & \textbf{0.926} & \textbf{*0.04} \\ \hline

\textbf{{[}Comedy=MaxPrefer{]}} & \textbf{1.299} & \textbf{0.521} & \textbf{3.665} & \textbf{1.321} & \textbf{10.166} & \textbf{*0.013} \\ 
{[}Comedy=MinPefer{]} & 0.055 & 0.695 & 1.057 & 0.271 & 4.124 & 0.936 \\ \hline

\textbf{{[}Crime=MaxPrefer{]}} & \textbf{1.295} & \textbf{0.469} & \textbf{3.65} & \textbf{1.456} & \textbf{9.151} & \textbf{*0.006} \\ 
{[}Crime=MinPefer{]} & 0.339 & 0.473 & 1.404 & 0.555 & 3.55 & 0.474 \\ \hline

\textbf{{[}Drama=MaxPrefer{]}} & \textbf{-0.905} & \textbf{0.428} & \textbf{0.405} & \textbf{0.175} & \textbf{0.935} & \textbf{*0.034} \\ 
{[}Drama=MinPefer{]} & -0.163 & 0.427 & 0.849 & 0.368 & 1.96 & 0.702 \\ \hline

{[}Documentary=MaxPrefer{]} & 0.237 & 0.414 & 1.267 & 0.563 & 2.854 & 0.567 \\ 
{[}Documentary=MinPefer{]} & -0.459 & 0.403 & 0.632 & 0.287 & 1.394 & 0.255 \\ \hline

{[}Fantasy=MaxPrefer{]} & -0.818 & 0.655 & 0.441 & 0.122 & 1.594 & 0.212 \\ 
{[}Fantasy=MinPefer{]} & -0.612 & 0.649 & 0.542 & 0.152 & 1.937 & 0.346 \\ \hline

{[}FlimNoir=MaxPrefer{]} & 0.017 & 0.423 & 1.017 & 0.444 & 2.329 & 0.968 \\ 
{[}FlimNoir=MinPefer{]} & -0.319 & 0.362 & 0.727 & 0.358 & 1.479 & 0.379 \\ \hline

\textbf{{[}Horror=MaxPrefer{]}} & \textbf{0.673} & \textbf{0.302} & \textbf{1.96} & \textbf{1.084} & \textbf{3.544} & \textbf{*0.026} \\ 
\textbf{{[}Horror=MinPefer{]}} & \textbf{0.898} & \textbf{0.298} & \textbf{2.454} & \textbf{1.369} & \textbf{4.4} & \textbf{*0.003} \\ \hline

{[}History=MaxPrefer{]} & -1.576 & 1.058 & 0.207 & 0.026 & 1.644 & 0.136 \\ 
{[}History=MinPrefer{]} & -0.786 & 1.323 & 0.456 & 0.034 & 6.095 & 0.552 \\ \hline

{[}Mystery=MaxPrefer{]} & -0.787 & 0.5 & 0.455 & 0.171 & 1.213 & 0.116 \\ 
{[}Mystery=MinPefer{]} & -0.317 & 0.52 & 0.728 & 0.263 & 2.019 & 0.542 \\ \hline

{[}Musical=MaxPrefer{]} & -0.148 & 0.389 & 0.862 & 0.402 & 1.848 & 0.703 \\ 
{[}Musical=MinPefer{]} & 0.224 & 0.351 & 1.251 & 0.628 & 2.49 & 0.524 \\ \hline

\textbf{{[}Romance=MaxPrefer{]}} & \textbf{-0.927} & \textbf{0.425} & \textbf{0.396} & \textbf{0.172} & \textbf{0.91} & \textbf{*0.029} \\ 
{[}Romance=MinPefer{]} & 0.017 & 0.423 & 1.017 & 0.444 & 2.331 & 0.967 \\ \hline

{[}SciFi=MaxPrefer{]} & 0.303 & 0.572 & 1.354 & 0.441 & 4.154 & 0.597 \\ 
{[}SciFi=MinPefer{]} & 0.221 & 0.604 & 1.247 & 0.382 & 4.071 & 0.714 \\ \hline

{[}Sport=MaxPrefer{]} & -1.753 & 1.173 & 0.173 & 0.017 & 1.728 & 0.135 \\ 
{[}Sport=MinPrefer{]} & -0.562 & 1.257 & 0.57 & 0.048 & 6.704 & 0.655 \\ \hline

{[}Thriller=MaxPrefer{]} & -0.72 & 0.551 & 0.487 & 0.165 & 1.434 & 0.192 \\ 
{[}Thriller=MinPefer{]} & -0.809 & 0.571 & 0.445 & 0.146 & 1.363 & 0.156 \\ \hline

\textbf{{[}War=MaxPrefer{]}} & \textbf{1.582} & \textbf{0.412} & \textbf{4.867} & \textbf{2.17} & \textbf{10.918} & \textbf{*0.000} \\ 
{[}War=MinPefer{]} & 0.369 & 0.404 & 1.447 & 0.656 & 3.194 & 0.36 \\ \hline

{[}Western=MaxPrefer{]} & 0.551 & 0.415 & 1.735 & 0.77 & 3.909 & 0.184 \\ 
{[}Western=MinPefer{]} & 0.057 & 0.357 & 1.058 & 0.526 & 2.129 & 0.874 \\ \hline

\multicolumn{7}{l}{{\begin{tabular}[c]{@{}l@{}}Goodness of fit: Chi-square = 532.125, df = 585 (p-value = 0.942)\\ 
a. ***The reference class variable's value for all genres is \textbf{No} \\ 
b. *Considered p-value <0.05 and marked the corresponding row \end{tabular}}} \\
\end{longtable}
}

We again conducted a binary logistic regression analysis to identify the influential genres and their impact on typical female movie preferences. The results are presented in Table \ref{tab:femaleLG}. As shown, the findings indicate a significant genre impact that is opposite to the results in Table \ref{tab:maleLG}. According to Table \ref{tab:femaleLG}, female viewers show a positive significant preference for \textbf{animation, drama, family,} and \textbf{romance} genres (p<0.05). Conversely, they exhibit a significant negative preference for \textbf{action, adventure, comedy, crime, horror}, and \textbf{war} genres (p<0.05). 

Table \ref{tab:femaleLG} highlights that females are 2.78 times more likely to prefer \textbf{animation} movies than males (odds = 2.783, p < 0.05), whereas their likelihood of choosing \textbf{action} and \textbf{adventure} movies are 81\% and 74\% lower compared to male viewers (action: odds = 0.185, p < 0.05; adventure: odds = 0.26, p < 0.05). Similarly, women show a strong preference for the \textbf{family} genre (maximum preference: odds = 6.676, p < 0.05; minimum preference: odds = 5.257, p < 0.05), while their preference for \textbf{comedy} and \textbf{crime} genres is significantly lower—by 73\%—compared to men (comedy: odds = 0.273, p < 0.05; crime: odds = 0.274, p < 0.05).

Additionally, women are 2.47 times more likely to prefer \textbf{drama} (odds = 2.471, p < 0.05) and 2.53 times more likely to choose \textbf{romance} (odds = 2.526, p < 0.05) compared to men. However, females are significantly less inclined to select \textbf{horror} (maximum preference: odds = 0.51, p < 0.05; minimum preference: odds = 0.407, p < 0.05) and \textbf{war} genres (maximum preference: odds = 0.205, p < 0.05) compared to male viewers. 

Moreover, the Chi-square and p-value results from the goodness-of-fit test indicate that the model fits the data well, ensuring the reliability of these findings.  

{
\small
\begin{longtable}{lcccccc}
\caption{\newline Multivariate Binary Logistic Regression Defining Influential Movie Genres of Females for Movie Preferences}
\label{tab:femaleLG} \\
\toprule
\multicolumn{1}{c}{\multirow{2}{*}{\textbf{Variables}}} & \multirow{2}{*}{\textbf{\begin{tabular}[c]{@{}c@{}}Logistic \\ Coefficient\end{tabular}}} & \multirow{2}{*}{\textbf{\begin{tabular}[c]{@{}c@{}}Standard \\ Error\end{tabular}}} & \multirow{2}{*}{\textbf{\begin{tabular}[c]{@{}c@{}}Odds \\ Ratio\end{tabular}}} & \multicolumn{2}{c}{\textbf{95\% CI}} & \multirow{2}{*}{\textbf{p-value$^b$}} \\ \cline{5-6}
\multicolumn{1}{c}{} &  &  &  & \textit{\textbf{\begin{tabular}[c]{@{}c@{}}Lower Bound\end{tabular}}} & \textit{\textbf{\begin{tabular}[c]{@{}c@{}}Upper Bound\end{tabular}}} &  \\ 
\midrule
\endfirsthead
\caption[]{\newline Multivariate Binary Logistic Regression Defining Influential Movie Genres of Females for Movie Preferences (continued)} \\
\toprule
\multicolumn{1}{c}{\multirow{2}{*}{\textbf{Variables}}} & \multirow{2}{*}{\textbf{\begin{tabular}[c]{@{}c@{}}Logistic \\ Coefficient\end{tabular}}} & \multirow{2}{*}{\textbf{\begin{tabular}[c]{@{}c@{}}Standard \\ Error\end{tabular}}} & \multirow{2}{*}{\textbf{\begin{tabular}[c]{@{}c@{}}Odds \\ Ratio\end{tabular}}} & \multicolumn{2}{c}{\textbf{95\% CI}} & \multirow{2}{*}{\textbf{p-value$^b$}} \\ \cline{5-6}
\multicolumn{1}{c}{} &  &  &  & \textit{\textbf{\begin{tabular}[c]{@{}c@{}}Lower Bound\end{tabular}}} & \textit{\textbf{\begin{tabular}[c]{@{}c@{}}Upper Bound\end{tabular}}} &  \\ 
\midrule
\endhead
\bottomrule
\multicolumn{7}{r}{{Continued on next page}} \\ 
\endfoot
\endlastfoot

\textbf{{[}Action=MaxPrefer{]}} & \textbf{-1.69} & \textbf{0.49} & \textbf{0.185} & \textbf{0.071} & \textbf{0.482} & \textbf{*0.001} \\
{[}Action=MinPefer{]} & -0.869 & 0.529 & 0.419 & 0.149 & 1.182 & 0.1 \\ 
{[}Action=No{]} & \multicolumn{6}{l}{***$Ref^a$} \\
\hline

\textbf{{[}Adventure=MaxPrefer{]}} & \textbf{-1.348} & \textbf{0.541} & \textbf{0.26} & \textbf{0.09} & \textbf{0.75} & \textbf{*0.013} \\
{[}Adventure=MinPefer{]} & -0.555 & 1.001 & 0.574 & 0.081 & 4.088 & 0.58 \\
\hline

\textbf{{[}Animation=MaxPrefer{]}} & \textbf{1.024} & \textbf{0.388} & \textbf{2.783} & \textbf{1.301} & \textbf{5.954} & \textbf{*0.008} \\
{[}Animation=MinPefer{]} & 0.687 & 0.393 & 1.987 & 0.92 & 4.293 & 0.081 \\
\hline

{[}Biography=MaxPrefer{]} & -0.506 & 0.885 & 0.603 & 0.106 & 3.416 & 0.567 \\
{[}Biography=MinPrefer{]} & -1.036 & 0.861 & 0.355 & 0.066 & 1.919 & 0.229 \\
\hline

\textbf{{[}Family=MaxPrefer{]}} & \textbf{1.899} & \textbf{0.821} & \textbf{6.676} & \textbf{1.335} & \textbf{33.381} & \textbf{*0.021} \\
\textbf{{[}Family=MinPefer{]}} & \textbf{1.66} & \textbf{0.808} & \textbf{5.257} & \textbf{1.08} & \textbf{25.593} & \textbf{*0.04} \\
\hline

\textbf{{[}Comedy=MaxPrefer{]}} & \textbf{-1.299} & \textbf{0.521} & \textbf{0.273} & \textbf{0.098} & \textbf{0.757} & \textbf{*0.013} \\
{[}Comedy=MinPefer{]} & -0.055 & 0.695 & 0.946 & 0.242 & 3.692 & 0.936 \\
\hline

\textbf{{[}Crime=MaxPrefer{]}} & \textbf{-1.295} & \textbf{0.469} & \textbf{0.274} & \textbf{0.109} & \textbf{0.687} & \textbf{*0.006} \\
{[}Crime=MinPefer{]} & -0.339 & 0.473 & 0.712 & 0.282 & 1.801 & 0.474 \\
\hline

\textbf{{[}Drama=MaxPrefer{]}} & \textbf{0.905} & \textbf{0.428} & \textbf{2.471} & \textbf{1.069} & \textbf{5.713} & \textbf{*0.034} \\
{[}Drama=MinPefer{]} & 0.163 & 0.427 & 1.177 & 0.51 & 2.717 & 0.702 \\
\hline

{[}Documentary=MaxPrefer{]} & -0.237 & 0.414 & 0.789 & 0.35 & 1.777 & 0.567 \\
{[}Documentary=MinPefer{]} & 0.459 & 0.403 & 1.582 & 0.718 & 3.489 & 0.255 \\
\hline

{[}Fantasy=MaxPrefer{]} & 0.818 & 0.655 & 2.266 & 0.627 & 8.188 & 0.212 \\
{[}Fantasy=MinPefer{]} & 0.612 & 0.649 & 1.844 & 0.516 & 6.585 & 0.346 \\
\hline

{[}FlimNoir=MaxPrefer{]} & -0.017 & 0.423 & 0.983 & 0.429 & 2.251 & 0.968 \\
{[}FlimNoir=MinPefer{]} & 0.319 & 0.362 & 1.375 & 0.676 & 2.796 & 0.379 \\
\hline

\textbf{{[}Horror=MaxPrefer{]}} & \textbf{-0.673} & \textbf{0.302} & \textbf{0.51} & \textbf{0.282} & \textbf{0.922} & \textbf{*0.026} \\
\textbf{{[}Horror=MinPefer{]}} & \textbf{-0.898} & \textbf{0.298} & \textbf{0.407} & \textbf{0.227} & \textbf{0.73} & \textbf{*0.003} \\
\hline

{[}History=MaxPrefer{]} & 1.576 & 1.058 & 4.837 & 0.608 & 38.461 & 0.136 \\
{[}History=MinPrefer{]} & 0.786 & 1.323 & 2.195 & 0.164 & 29.364 & 0.552 \\
\hline

{[}Mystery=MaxPrefer{]} & 0.787 & 0.5 & 2.197 & 0.824 & 5.856 & 0.116 \\
{[}Mystery=MinPefer{]} & 0.317 & 0.52 & 1.373 & 0.495 & 3.807 & 0.542 \\
\hline

{[}Musical=MaxPrefer{]} & 0.148 & 0.389 & 1.16 & 0.541 & 2.486 & 0.703 \\
{[}Musical=MinPefer{]} & -0.224 & 0.351 & 0.799 & 0.402 & 1.591 & 0.524 \\
\hline

\textbf{{[}Romance=MaxPrefer{]}} & \textbf{0.927} & \textbf{0.425} & \textbf{2.526} & \textbf{1.099} & \textbf{5.808} & \textbf{*0.029} \\
{[}Romance=MinPefer{]} & -0.017 & 0.423 & 0.983 & 0.429 & 2.252 & 0.967 \\
\hline

{[}SciFi=MaxPrefer{]} & -0.303 & 0.572 & 0.739 & 0.241 & 2.267 & 0.597 \\
{[}SciFi=MinPefer{]} & -0.221 & 0.604 & 0.802 & 0.246 & 2.617 & 0.714 \\
\hline

{[}Sport=MaxPrefer{]} & 1.753 & 1.173 & 5.77 & 0.579 & 57.512 & 0.135 \\
{[}Sport=MinPrefer{]} & 0.562 & 1.257 & 1.754 & 0.149 & 20.625 & 0.655 \\
\hline

{[}Thriller=MaxPrefer{]} & 0.72 & 0.551 & 2.054 & 0.697 & 6.052 & 0.192 \\
{[}Thriller=MinPefer{]} & 0.809 & 0.571 & 2.245 & 0.734 & 6.869 & 0.156 \\
\hline

\textbf{{[}War=MaxPrefer{]}} & \textbf{-1.582} & \textbf{0.412} & \textbf{0.205} & \textbf{0.092} & \textbf{0.461} & \textbf{*0.000} \\
{[}War=MinPefer{]} & -0.369 & 0.404 & 0.691 & 0.313 & 1.526 & 0.36 \\
\hline

{[}Western=MaxPrefer{]} & -0.551 & 0.415 & 0.576 & 0.256 & 1.299 & 0.184 \\
{[}Western=MinPefer{]} & -0.057 & 0.357 & 0.945 & 0.47 & 1.901 & 0.874 \\ 
\hline

\multicolumn{7}{l}{{\begin{tabular}[c]{@{}l@{}}Goodness of fit: Chi-square = 532.125, df = 585 (p-value = 0.942)\\ 
a. ***The reference class variable's value for all genres is \textbf{No} \\ 
b. *Considered p-value <0.05 and marked the corresponding row \end{tabular}}} \\
\end{longtable}
}

\subsubsection{\textbf{Discussion of RQ1}}

This user study investigated the gender stereotypes for movie preferences of the viewers in the context of actually favorable genres at the time of movie selection for watching. Usually, a viewer selects genres that align with their fondness and selects a movie that belongs to the corresponding genres \citep{steck2018calibrated}. Our analysis portrays that viewers' gender is strongly associated with movie genres. As per the regression results (Table \ref{tab:maleLG} \& \ref{tab:femaleLG}), it can be deduced that each or a combination of action, adventure, comedy, crime, horror \& war genre-related movies have a pragmatic impact on male viewers to select a movie for their enjoyment while also have a negative effect on females. Conversely, animation, family, romance \& drama genre-related movies are positively affiliated with the female viewer's movie selection motive and negatively associated with the male viewer's motive for the movie selection. Additionally, it can be claimed that the higher preference scores for action, adventure, comedy, crime, horror \& war genre-related movies are the positive predictor for the male gender and the negative predictor for the female gender with the lower preference scores. On the other side, it can also be stated that the female gender can be predicted by the superior choice indicators of romance, drama, family \& animation genre-related movies, and the male gender can be assumed with the lower choice indicators of the same set of movies' genres. Furthermore, our findings also mostly aligned with our hypotheses and the existing literature with some additions and support the media content, social theory \& biological accounts as the source of the GSs for movie preferences \citep{wuhr2017tears, infortuna2021inner, palomba2020consumer, martin2019you, procyshyn2020experimental}. Table \ref{tab:hys} summarises our findings with the hypotheses mentioned in section \ref{sec:GSD}. 

\begin{table}[]
\centering
\caption{Findings Summary with Hypotheses.}
\label{tab:hys}
\begin{tabular}{ccl}
\hline
\multirow{2}{*}{\textbf{Genre}} & \multirow{2}{*}{\textbf{Results}} & \multicolumn{1}{c}{\multirow{2}{*}{\textbf{Summary}}} \\
 &  & \multicolumn{1}{c}{} \\
\hline
\multicolumn{3}{l}{\textit{\begin{tabular}[c]{@{}l@{}}H1: Females prefer movies centered on social relationships more than males, especially those affiliated with romance, \\drama, and family genres.\end{tabular}}} \\
\hline
Romance & Supported & \multirow{4}{*}{\begin{tabular}[c]{@{}l@{}}Females prefer movies that belong to romance, drama, family, including \\ animation genres on average 5 times more than males (based on odd ratios \& \\ p-values). Because these genres are centred on social relationships. \end{tabular}} \\
Drama & Supported &  \\
Family & Supported &  \\
Animation & \textbf{Additional Finding} &  \\
\hline
\multicolumn{3}{l}{\textit{\begin{tabular}[c]{@{}l@{}}H2: Males prefer movies centered on aggressive conflicts \& dominant themes more than females, particularly in genres \\ such as horror, action, adventure, fantasy, crime, and war.\end{tabular}}} \\
\hline
Horror & Supported & \multirow{7}{*}{\begin{tabular}[c]{@{}l@{}}Males prefer movies that belong to action, adventure, crime, horror, and war, \\ including comedy genres as these genres centered on aggressive conflict \& \\ dominant motives approximately 4 times higher than females (based on odd \\ ratios \& p-values).\end{tabular}} \\
Action & Supported &  \\
Adventure & Supported &  \\
Fantasy & Not Supported &  \\
Crime & Supported &  \\
War & Supported &  \\
Comedy & \textbf{Additional Finding} & \\
\hline
\end{tabular}
\end{table}

\subsection{Methods and results for RQ2}
\label{sec:RQ2}

\begin{itemize}
    \item[] \textbf{RQ2: } \textit{Does the gender stereotype exist in the movie recommender system?}
\end{itemize}

We experimented on publicly available movie recommender datasets with some inference algorithms to find the answer to our second research question. We validated our findings concerning the state-of-the-art performance indicators of the algorithms, which are detailed mentioned in the consecutive subsections as follows:  

\subsubsection{\textbf{Data Preparation}} 

To thoroughly investigate gender stereotypes in movie recommender systems, we chose two publicly accessible datasets that contain gender information: MovieLens and Yahoo!Movie\footref{YM}. Specifically, we used the MovieLens 1 million (ML1M)\footref{ML} dataset for MovieLens. Both datasets provide ratings on a scale of 1 to 5 as the explicit interaction data and include gender as a binary attribute, categorizing users as either male or female.

\textbf{Dataset: MovieLens 1M}. The MovieLens 1M dataset contains 72\% male and 28\% female data. Out of 18 genres, all the movies in this dataset belong to either one or more than one genre. It contains a total of 1000209 ratings. In this dataset, there was a genre named "Children's," which we treated as the "Family" genre to align with IMDb.

\textbf{Dataset: Yahoo!Movie}. The Yahoo!Movie dataset contains 71\% male and 29\% female data. We had to process the dataset's unevenness before using it in the experiments. We identified a total of 5 types of unevenness existence in the dataset. The dataset includes a total of 29 genres with some redundancy. It has $"Miscellaneous"$ (movie id: 1800022403) \& $"Features"$ (movie id: 1802828612) as genres. However, these genres are not valid. To resolve this unevenness, we manually checked the IMDb website with the movie titles and replaced the invalid genres with the corresponding correct genres in the dataset. Secondly, there were some similar genres paired, such as Gangster \& Crime, and Music, Musical \& Performing Art, that were converted into one genre, such as Crime and Music, respectively. Also, we converted the Kids \& Family genre pair into the Family genre as the genre pair would encompass movies that cater to children while providing content that adults can enjoy. For example, many successful animated films from studios like Pixar and Disney already offer multi-layered stories that entertaining viewers of all ages. Furthermore, we modified the Suspense \& Thriller genre pair into the Thriller genre because both genres are designed to evoke excitement, tension, and anticipation in the audience. The third category of unevenness is the dataset contains some videos (for example, movie id: 1800219584, 1800149736 and many more) that are not movies under the genre "$Adult \; Audience$"; we also omitted the genre with the associated items for our experiments. Table \ref{tab:yahooGen} shows the complete list of genre conversions. 

\begin{table}[ht]
\centering
\caption{Uneven Genre Conversion of Yahoo!Movie}
\label{tab:yahooGen}
\begin{tabular}{cccc}
\hline
\textbf{Genre in Dataset} & \textbf{Converted Genre} & \textbf{Genre in Dataset} & \textbf{Converted Genre} \\
\hline
Miscellaneous and Features & Actual Genres List from IMDb & Suspense, Thriller & Thriller \\
Music, Musical and Performing Art & Musical & Kids, Family & Family \\
Adult Audience and Delete & (-) removed & Gangster, Crime & Crime\\
\hline
\end{tabular}
\end{table}

Furthermore, some movies contain no information about genres, which we retrieved from the IMDb. For example, movieId: 1808404630, title: Jet Lag (2003), belongs to the "Comedy|Romance" genres, but the corresponding movie item contains no genre information in the dataset. So, we replaced the actual genres with the null genre value in the dataset. By following the same approach, we corrected more than 300+ data in this category of uneven data. Moreover, the fifth category of unevenness is the dataset containing some TV series data. We also erased those data (for example, movie id: 1808483376, 1808523840, 1808407101, and much more) as these are not movies. After mitigating the unevenness, the final dataset contains 7637 users and 9293 movies with 207307 ratings \& 19 genres. All the required statistics of the MovieLens 1M and Yahoo!Movie datasets for our experiments are listed in Table \ref{tab:dataset}.  


\begin{table}[ht]
\renewcommand{\arraystretch}{1.2}
\centering
\caption{Datasets Description}
\label{tab:dataset}
\begin{tabular}{cccccccc}
\hline
Datasets & Users & Male & Female & Movies & Ratings & Movie Genres & Density (\%) \\ \hline
MovieLens 1M & 6040  & 4331 & 1709   & 3952   & 1000209 & 18 & 4.19 \\ 
Yahoo!Movie & 7637  & 5432 & 2205   & 9293   & 207307 & 19 & 1.41 \\ \hline
\end{tabular}
\end{table}

\subsubsection{\textbf{Inference Algorithm}} To conduct the experiments, logistic regression (LR), support vector machine (SVM), AdaBoost, and XGBoost classifiers were chosen as the gender inference algorithm. \textbf{Logistic regression (LR)} is a well-known supervised machine learning (ML) technique utilized primarily for binary classification tasks. It predicts the likelihood of an input belonging to a specific class by employing a logistic (sigmoid) function. This function maps the input features to a probability value ranging between 0 and 1, indicating the probability of the positive class. Conversely, the \textbf{support vector machine (SVM)} is another ML algorithm commonly used for classification and regression analysis. It works by finding the hyperplane that maximally separates the classes in a given dataset. The hyperplane is chosen to maximize the margin between the two classes. In the previous literature \citep{base01_2021Gender, weinsberg2012blurme, feng2014tags}, LR and SVM mainly applied as a classifier to infer the users' gender. 

On the other side, \textbf{XGBoost} denotes Extreme Gradient Boosting, a supervised ensemble ML classifier. It's rooted in the concept of decision trees within a gradient-boosting framework. XGBoost tackles various tasks, such as regression, classification, ranking, and user-defined prediction. As a fully modular tree-boosting algorithm, it is extensively utilized in ML for its efficiency in tackling real-world problems with minimal resources. Further, \textbf{Adaptive Boosting (AdaBoost)} is a renowned ensemble learning technique used in ML. It trains weak classifiers iteratively, usually simple models like decision trees, and assigns higher weights to misclassified instances in each iteration. These weak classifiers are then combined to form a robust classifier through a weighted sum, where classifiers with higher accuracy contribute more to the final prediction. AdaBoost is adequate for binary and multiclass classification tasks and performs well with weak learners. It improves the model's overall performance by focusing on difficult-to-classify instances in each iteration, making it robust against overfitting and suitable for various applications.

\subsubsection{\textbf{Evaluation Parameters}}

Typically, each classifier produces a confusion matrix that displays the number of correctly and incorrectly classified instances. Various statistical metrics are then derived from this confusion matrix to assess the classifier's performance. In the matrix, correctly classified instances are represented as $TP$ for "True Positive" and $TN$ for "True Negative", while incorrectly classified instances are denoted as $FP$ for "False Positive" and $FN$ for "False Negative". Specifically, $TP$ indicates the correctly identified positive cases, while $FP$ represents the positive cases incorrectly classified by the algorithm. Similarly, $TN$ corresponds to correctly identified negative cases, and $FN$ indicates the negative cases incorrectly classified. To evaluate the classifier's effectiveness, we use the following parameters.

\textbf{Accuracy.} The accuracy parameter measures the ratio of correctly classified classes in the test sample by the classifier. Equation \ref{eq:accuracy} shows the mathematical formulation of accuracy calculation. For the rest of this study, we referred to the classifier accuracy as 'overall accuracy' to distinguish it from specific class accuracy.  

\begin{equation}
\label{eq:accuracy}
    Accuracy = \frac{TP+TN}{TP+FP+TN+FN}
\end{equation} 

\textbf{Accuracy for a specific class.} This parameter measures the proportion of correctly predicted samples for a class out of all actual samples of that particular class. In our experiment, we have two classes where the male user group is treated as a positive class, and the female user group is assumed as a negative class. For better understanding, we will denote the positive class as "Male" and the negative class as "Female". This metric helps assess how well the model performs for each class individually, focusing only on the actual sample set of the respective class. The formulas for calculating the accuracy for the specific class are presented as follows: 

\begin{equation}
    \begin{matrix}
        Accuracy_{Male} = \frac{TP}{TP+FN}
        \\\\
        Accuracy_{Female} = \frac{TN}{TN+FP}
    \end{matrix}
\end{equation}

\textbf{Precision.} It defines the proportion of correctly classified class instances among all the cases predicted as correct instances by the classifier. Usually, precision reflects the classifier's accuracy in positive predictions and is the second most widely used evaluation metric \citep{afsar2022reinforcement}. Equation \ref{eq:precision} denotes how the precision is calculated.

\begin{equation}
\label{eq:precision}
    Precision = \frac{TP}{TP+FP}
\end{equation} 

\textbf{Recall.} Recall is also called the sensitivity of the evaluated algorithm \citep{hasan2019item}. It indicates the proportion of correctly classified positive instances out of all positive instances by the classifier. Recall is measured by using the Equation \ref{eq:recall}. 

\begin{equation}
\label{eq:recall}
    Recall = \frac{TP}{TP+FN}
\end{equation} 

\textbf{F-measure.} F-measure is also denoted as f-score and combines the precision and recall metrics into a single metric, offering a balanced assessment of both aspects \citep{roy2023item}. F-measure is calculated by the weighted harmonic mean of the precision and recall as the following Equation \ref{eq:f1}.

\begin{equation}
\label{eq:f1}
    F_{measure} = \frac{2*percision*recall}{percision+recall}
\end{equation}



\textbf{Area Under Curve (AUC). } The Area Under the Curve denotes the area under the Receiver Operating Characteristic (ROC) curve that illustrates the diagnostic ability of a binary classifier system as its discrimination threshold is varied. AUC quantifies the performance of a classifier across all thresholds by calculating the area under the ROC curve. It offers a single numerical value representing the classifier's capability to differentiate between positive and negative classes. A higher AUC signifies superior classifier performance, with 1 denoting perfection and 0.5 indicating performance akin to random guessing \citep{base01_2021Gender}. AUC is statistically a better choice for evaluating the classifier performance than the accuracy metric \citep{zhu2017empirical}.

\subsubsection{\textbf{Experiments: Comparison Between Inference Algorithms}} 


To compare the performance of gender inference algorithms, we conducted our experiments in two folds. Initially, four classifier algorithms were selected for the first fold to perform preliminary experiments. Based on the results, we identified the top two classifiers that outperformed the others and then conducted rigorous experiments using five evaluation parameters. We applied L2-normalization\footnote{https://scikit-learn.org/stable/modules/generated/sklearn.preprocessing.normalize.html} in each fold to standardize the feature vector before feeding it into the classifiers. To handle the imbalance in sample sizes between male and female classes, we applied the 'class weight' parameter to assign greater penalties to misclassifications of the minority class compared to the majority class. In both datasets of this study, the male class represents the majority, while the female class constitutes the minority. The 'class weight' parameter adjusts the weights to encourage the classifier to pay greater attention to the minority class, thereby balancing the influence of each class during training.

In the first fold, we randomly split our dataset into an 80:20 ratio, with 80\% of the data used for training and the remaining 20\% allocated for testing the classifier's performance. This 80:20 split was chosen because it is suitable for the initial assessment of model performance before employing more rigorous validation techniques and is well-suited for quick evaluations. We used AUC, overall classifier accuracy, and accuracy for specific classes as the evaluation metrics for this experiment fold. Overall classifier accuracy measures the total number of correctly predicted instances by the classifier, and AUC assesses the classifier's ability to distinguish between true positive and true negative instances. Moreover, accuracy for specific classes calculates the proportion of correctly predicted instances for each respective class through the classifier. 

Additionally, we conducted experiments with four different input sets. Two sets contained only user-item interaction data from the MovieLens 1M and Yahoo!Movie datasets, while the other two sets incorporated gender stereotypes (GSs) data along with the user-item interaction data for both datasets. For the logistic regression classifier, the regularization strength was set to 1, the L2 norm was selected as the ridge regularization technique, and 1000 iterations were chosen as the maximum limit for the solver to converge. For AdaBoost, we set the number of weak learners in the ensemble-based classifier to 50. For XGBoost, the objective function "multi" was applied for binary classification, while for SVM, we selected a linear kernel to conduct the experiments. Table \ref{tab:performance_classifier} presents the results of our first-fold experiments. We used "rating data" and "interaction data" interchangeably for this study. For this reason, the input sets in Table \ref{tab:performance_classifier} containing only user-item interaction information are labeled as \textbf{\textit{"Rating Data"}}, while the input sets that included user-item interaction with GSs information are labeled as \textbf{\textit{"Rating Data with GS Association"}}.

\begin{table}[h!]
\renewcommand{\arraystretch}{1.2}
\centering
\caption{Performance Comparison of Different Classifiers }
\label{tab:performance_classifier}
\begin{tabular}{cccccc}
\hline
\multirow{2}{*}{Classifiers} & \multirow{2}{*}{\begin{tabular}[c]{@{}c@{}}Evaluation \\ Parameter\end{tabular}} & \multicolumn{2}{c}{MovieLens 1M} & \multicolumn{2}{c}{Yahoo!Movie} \\ \cline{3-6} 
 &  & \multicolumn{1}{c}{\begin{tabular}[c]{@{}c@{}}Rating Data\\ (\textbf{\%})\end{tabular}} & \begin{tabular}[c]{@{}c@{}}Rating Data with \\GS Association (\textbf{\%})\end{tabular} & \multicolumn{1}{c}{\begin{tabular}[c]{@{}c@{}}Rating Data\\ (\textbf{\%})\end{tabular}} & \begin{tabular}[c]{@{}c@{}}Rating Data with \\ GS Association (\textbf{\%})\end{tabular} \\ \hline
 
\multirow{4}{*} {\begin{tabular}[c]{@{}c@{}}Logistic \\ Regression\end{tabular}} 
& AUC & 
\multicolumn{1}{c}{87.70} & \underline{89.07} & \multicolumn{1}{c}{80.99}  & \underline{84.80} \\ 
& Accuracy & 
\multicolumn{1}{c}{79.50} & \underline{82.65} & \multicolumn{1}{c}{74.93} & \underline{76.28} \\ 
& Accuracy$_{MaleClass}$ & 
78.80 & \underline{81.20} & 77.10 & \underline{77.91} \\
& Accuracy$_{FemaleClass}$ & 
81.30 & \underline{85.70} & 69.40 & \underline{72.06} \\
\hline

\multirow{4}{*}{XGBoost} 
& AUC & 
82.30 & 85.51 & 77.69 & 80.98 \\ 
& Accuracy & 
77.61 & 79.10 & 73.82 & 75.28 \\ 
& Accuracy$_{MaleClass}$ & 
84.30 & 86.20 & 78.50 & 80.73 \\
& Accuracy$_{FemaleClass}$ & 
60.30 & 62.30 & 59.10 & 61.21 \\
\hline

\multirow{4}{*}{AdaBoost} 
& AUC & 
78.20 & 81.31 & 75.59 & 76.41 \\ 
& Accuracy & 
69.71 & 75.99 & 69.37 & 72.19 \\ 
& Accuracy$_{MaleClass}$ & 
68.94 & 76.82 & 69.18 & 74.00 \\
& Accuracy$_{FemaleClass}$ & 
71.51 & 74.05 & 69.86 & 67.53 \\
\hline

\multirow{4}{*}{\textbf{SVM}} 
& AUC & 
87.80 & \textbf{89.18} & 80.84 & \textbf{85.03} \\ 
& Accuracy & 
80.04 & \textbf{82.54} & 74.86 & \textbf{76.26} \\ 
& Accuracy$_{MaleClass}$ & 
79.64 & \textbf{82.24} & 77.31 & \textbf{78.45} \\
& Accuracy$_{FemaleClass}$ & 
81.01 & \textbf{83.24} & 68.70 & \textbf{70.56} \\
\hline
\end{tabular}
\end{table} 

From Table \ref{tab:performance_classifier}, we observe that integrating gender stereotypes (GSs) with user-item interaction data positively impacts the inference algorithm, enabling it to detect gender more accurately than when using only interaction data. Usually, gender stereotypes can influence user behavior, such as preferences for specific genres of movies. By incorporating this data as part of the feature vector, the classifier can better identify patterns that enhance its ability to predict users' gender from their preferences. Integrating GSs with user-item interaction data also provides a more detailed representation of user profiles, making it easier for the classifier to differentiate users' genders more precisely. Moreover, as the user-item interaction matrix is often sparse, with many missing entries, GS integration adds supplementary information that provides valuable context to the classifiers. Consequently, this integration improved classifier performance for the MovieLens 1M dataset by approximately 3.36\% in overall accuracy, 2.27\% in AUC, 3.70\% in male class accuracy, and 2.80\% in female class accuracy compared to the classifier trained only on rating data. Similarly, for the Yahoo!Movie dataset, which has a higher degree of sparsity compared to MovieLens, the integration of GSs with rating data enhanced classifier performance by 1.75\% in accuracy, 3.08\% in AUC, 2.25\% in male class accuracy, and 1.08\% in female class accuracy when compared to results based solely on rating data.  

Furthermore, in previous studies \citep{base01_2021Gender, weinsberg2012blurme}, logistic regression (LR) performed better than other classifiers. However, in our experiment, SVM achieved the best performance (highlighted in bold in Table \ref{tab:performance_classifier}), while LR ranked second (underlined in Table \ref{tab:performance_classifier}) among all classifiers. Thus, we conducted the second-stage experiment using LR and SVM. For this experiment, we followed the same experimental settings as in \citep{base01_2021Gender}. We used ten-fold cross-validation with StratifiedKFold, where data from nine folds was used to train the classifier and data from the remaining one fold was used for evaluation. This ten-fold cross-validation was chosen because it provides a more stable performance metric and a robust estimate of model performance by reducing the variance associated with a single train-test split. It is also often used in conjunction with hyperparameter tuning to select the best model parameters. The hyperparameters were chosen from the training set using the GridSearchCV technique. The results of the second-fold experiment are summarized in Table \ref{tab:cv_LR} using accuracy, AUC, precision, recall, and F-score as evaluation metrics, with the standard deviation reported as follows.

\begin{table}[h!]
\renewcommand{\arraystretch}{1.2}
\centering
\caption{Average performance of logistic regression and SVM classifiers. The ± symbol indicates the standard deviation of the results across ten-fold cross-validation.}
\label{tab:cv_LR}
\begin{tabular}{cccccc}
\hline
\multirow{2}{*}{Classifiers} & \multirow{2}{*}{\begin{tabular}[c]{@{}c@{}}Evaluation \\ Parameter\end{tabular}} & \multicolumn{2}{c}{MovieLens 1M} & \multicolumn{2}{c}{Yahoo!Movie} \\ \cline{3-6} 
 &  & \multicolumn{1}{c}{Rating Data (\textbf{\%})} & \begin{tabular}[c]{@{}c@{}}Rating Data with \\ GS Association (\textbf{\%})\end{tabular} & \multicolumn{1}{c}{Rating Data (\textbf{\%})} & \begin{tabular}[c]{@{}c@{}}Rating Data with \\ GS Association (\textbf{\%})\end{tabular} \\ \hline
 
\multirow{5}{*} {\begin{tabular}[c]{@{}c@{}}Logistic \\ Regression\end{tabular}} 
& AUC & 
0.86 ± 0.016 & \underline{0.87 ± 0.015} & 0.81 ± 0.009 & \underline{0.83 ± 0.008} \\ 
& Accuracy & 
0.80 ± 0.017 & \underline{0.81 ± 0.011} & 0.75 ± 0.008 & \underline{0.78 ± 0.001} \\ 
& Precision & 
0.61 ± 0.000 & \underline{0.63 ± 0.000} & 0.55 ± 0.000 & \underline{0.57 ± 0.000} \\ 
& Recall & 
0.75 ± 0.000 & \underline{0.78 ± 0.000} & 0.71 ± 0.000 & \underline{0.74 ± 0.000} \\ 
& F-score & 
0.68 ± 0.000 & \underline{0.70 ± 0.000} & 0.62 ± 0.000 & \underline{0.65 ± 0.000} \\ 
\hline

\multirow{5}{*}{SVM} 
& AUC & 
0.86 ± 0.015 & \textbf{0.87 ± 0.014} & 0.81 ± 0.008 & \textbf{0.84 ± 0.008} \\ 
& Accuracy & 
0.80 ± 0.020 & \textbf{0.82 ± 0.013} & 0.76 ± 0.007 & \textbf{0.78 ± 0.009} \\ 
& Precision & 
0.62 ± 0.000 & \textbf{0.64 ± 0.000} & 0.57 ± 0.000 & \textbf{0.60 ± 0.000} \\ 
& Recall & 
0.76 ± 0.000 & \textbf{0.78 ± 0.000} & 0.69 ± 0.000 & \textbf{0.72 ± 0.000} \\ 
& F-score & 
0.68 ± 0.000 & \textbf{0.71 ± 0.000} & 0.62 ± 0.000 & \textbf{0.65 ± 0.000} \\ 
\hline
\end{tabular}
\end{table} 

Table \ref{tab:cv_LR} summarizes the performance of Logistic Regression and SVM classifiers across the MovieLens 1M and Yahoo!Movie datasets were evaluated using AUC, accuracy, precision, recall, and F-score metrics. The results demonstrate that integrating gender stereotype (GS) associations with user-item interaction data consistently improves classifier performance compared to only rating data. For instance, in the MovieLens 1M dataset, Logistic Regression achieves an AUC of 0.87 and accuracy of 0.81 with GS associations, compared to 0.86 and 0.80 without GS associations, respectively. Similarly, the SVM classifier shows an increase in AUC from 0.86 to 0.87 and accuracy from 0.80 to 0.82. The Yahoo!Movie dataset, which exhibits higher sparsity, also benefits from GS integration, with improvements observed across both classifiers. For example, the Logistic Regression classifier improves its AUC from 0.81 to 0.83 and accuracy from 0.75 to 0.78, while SVM shows corresponding improvements from 0.81 to 0.84 and 0.76 to 0.78. Furthermore, the classifiers' performance consistently improves when trained on the integration of GS and rating data, particularly in terms of recall, precision, and F-score evaluation metrics. A similar performance improvement pattern is observed in sparse datasets like Yahoo!Movie, highlighting the integration's ability to capture more detailed user preferences and enhance classification performance. These findings underscore the effectiveness of the classifiers in leveraging GS integration with interaction data to achieve more accurate gender inference.

Furthermore, after conducting the experiments and analyzing the results (Table \ref{tab:performance_classifier} and \ref{tab:cv_LR}), we conclude that the existing dataset of the movie recommender system contains gender-stereotypical data. This is evident from the fact that integrating GSs with interaction data improves the inference algorithm's performance by enhancing classification accuracy between classes and achieving better predictions across metrics such as precision, recall, F-score, and AUC. 

\subsection{Methods and results for RQ2.1}
\label{sec:RQ2.1}

\begin{itemize}
    \item[] \textbf{RQ2.1: } \textit{If GSs exist, what is the magnitude of the presence of gender stereotypes in the movie recommender system?}
\end{itemize}

\begin{figure*}[ht]
  \centering
  \includegraphics[width=.65\linewidth]{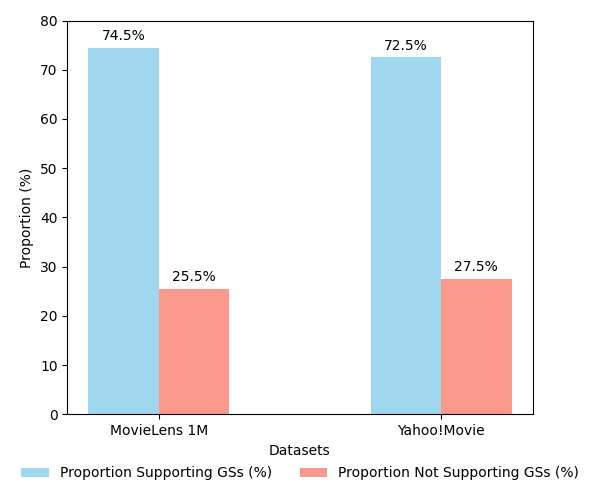}
  \caption{Extent of Gender Stereotypes in Movie Recommender Data}
  \label{fig:GSinMRS}
\end{figure*}

The previous section demonstrated the existence of gender stereotypes (GSs) in the movie recommender data, and this section aims to quantify the extent to which the data is gender-stereotypical. To achieve this, we applied Equation \ref{eq:GS_Count} in the experiments to calculate the proportion of users whose ratings align with or deviate from GSs. Figure \ref{fig:GSinMRS} highlights the degree of alignment between GSs and user rating behavior across two datasets: MovieLens 1M and Yahoo!Movie. 

Figure \ref{fig:GSinMRS} presents the extent of gender stereotypes (GSs) in the existing Movie Recommender datasets, specifically MovieLens 1M and Yahoo!Movie. The analysis reveals that in the MovieLens 1M dataset, 1542 users out of 6040 (25.5\%) have rating data that do not align with GSs, while 74.5\% exhibit rating patterns consistent with GSs. Similarly, in the Yahoo!Movie dataset, 2101 users out of 7637 (27.5\%) have rating data that do not match GSs, with 72.5\% supporting GS-aligned patterns. These findings demonstrate that a significant proportion of users in both datasets exhibit gender-stereotypical preferences in their movie ratings, with most users supporting GS-aligned behavior. Moreover, the degree of alignment between GSs and users' preference data provides valuable information for adversaries, enabling them to easily interpret users' gender with a high probability using classification algorithms.

\subsection{Important Findings of the Experiments}
\label{sec:GD}

After observing all the experimental results, we can conclude the following statements:

\begin{itemize}

    \item Based on user ratings, the existing data in movie recommender systems is highly aligned with gender stereotypes related to movie preferences.
    
    \item Ratings are an explicit form of user feedback. As such, movie ratings reflect the degree of users' preferences, as they are based on movies they have previously consumed. The genres of rated movies create an opportunity to measure the probability of a user's gender by leveraging knowledge of GSs associated with movie preferences.
    
    \item The findings from the first two observations (Table \ref{tab:performance_classifier} and \ref{tab:cv_LR}) highlight a critical issue that the existing MRS data is alarmingly sensitive concerning user privacy. An adversary equipped with the proposed threat model \ref{threatmodel} or any advanced threat model, combined with background knowledge such as GSs and users' interaction data, could exploit this vulnerability to infer private user attributes. 
    
    \item The experimental results (Figure \ref{fig:GSinMRS}) further demonstrate the association between users' public and private data. Generally, users watch movies and rate them based on their preferences. These interaction data, which align with GSs, can serve as interpretive knowledge to infer users' gender with high accuracy.  
    
    \item Additionally, based on users' demographic data, demographic biases, such as gender bias, may be introduced into the system for business purposes \citep{ekstrand2018exploring, melchiorre2021investigating, gomez2022provider}. For example, action movies are often promoted more to male audiences, while romantic animated movies are marketed with themes that attract female audiences. These practices can create gender-specific viewing patterns, reducing the variety of genres each gender gets to see and possibly restricting individual preferences.

    \item The potential for gender bias in the system can also result in issues such as filter bubbles or feedback loop effects. These phenomena confine users to repetitive, similar recommendations, thereby limiting their exposure to diverse content and reinforcing stereotypical behavior \citep{mansoury2020feedback}.

\end{itemize}

\section{Conclusions}
\label{sec:cons}

This study identified that movie recommender systems often contain gender stereotypes in their data, which raises significant privacy concerns. It is the first study to examine the privacy implications of these stereotypes, whereas previous research primarily focused on improving system performance using gender stereotypes. The study discovered that over 70\% of user feedback in public datasets reflected gender stereotypes in movie preferences, highlighting privacy issues. Future research could aim to protect users' privacy by reducing the link between their feedback and personal attributes. It could also look at ways to publish recommender system data that protects privacy without losing usefulness. Another area to explore is creating a measure to evaluate system-induced or data-oriented gender stereotypes in recommender systems. However, the study has some limitations. It only looked at gender as a private attribute and didn't consider other stereotypes like age. It focused on gender inference attacks, assuming that an adversary can access user ratings and knows about gender stereotypes in movie preferences. Lastly, the study focused solely on gender stereotypes in movie recommender data, which were propagated through the recommender algorithms and amplified in the system's output due to the recurring nature of recommendation processes.


\bibliographystyle{model1-num-names}


\begin{thebibliography}{10}
\expandafter\ifx\csname natexlab\endcsname\relax\def\natexlab#1{#1}\fi
\providecommand{\url}[1]{\texttt{#1}}
\providecommand{\path}[1]{#1}
\providecommand{\DOIprefix}{doi:}
\providecommand{\ArXivprefix}{arXiv:}
\providecommand{\URLprefix}{URL: }
\providecommand{\Pubmedprefix}{pmid:}
\providecommand{\doi}[1]{\href{http://dx.doi.org/#1}{\path{#1}}}
\providecommand{\Pubmed}[1]{\href{pmid:#1}{\path{#1}}}
\providecommand{\bibinfo}[2]{#2}
\ifx\xfnm\relax \def\xfnm[#1]{\unskip,\space#1}\fi
\bibitem[{Feng et~al.(2024)Feng, Yuan, Ye, Qian, and Ge}]{FENG2024103905}
\bibinfo{author}{L.~Feng}, \bibinfo{author}{H.~Yuan}, \bibinfo{author}{Q.~Ye}, \bibinfo{author}{Y.~Qian}, \bibinfo{author}{X.~Ge},
\newblock \bibinfo{title}{Exploring the impacts of a recommendation system on an e-platform based on consumers’ online behavioral data},
\newblock \bibinfo{journal}{Information \& Management} \bibinfo{volume}{61} (\bibinfo{year}{2024}) \bibinfo{pages}{103905}.
\bibitem[{Tian et~al.(2023)Tian, Shi, and Li}]{TIAN2023103815}
\bibinfo{author}{J.~Tian}, \bibinfo{author}{X.~Shi}, \bibinfo{author}{M.~Li},
\newblock \bibinfo{title}{Price-aware matrix factorization model for personalized recommendations},
\newblock \bibinfo{journal}{Information \& Management} \bibinfo{volume}{60} (\bibinfo{year}{2023}) \bibinfo{pages}{103815}.
\bibitem[{Shi et~al.(2023)Shi, Li, Ding, and Bu}]{shi2023selection}
\bibinfo{author}{L.~Shi}, \bibinfo{author}{S.~Li}, \bibinfo{author}{X.~Ding}, \bibinfo{author}{Z.~Bu},
\newblock \bibinfo{title}{Selection bias mitigation in recommender system using uninteresting items based on temporal visibility},
\newblock \bibinfo{journal}{Expert Systems with Applications} \bibinfo{volume}{213} (\bibinfo{year}{2023}) \bibinfo{pages}{118932}.
\bibitem[{Wu et~al.(2022)Wu, Ma, Zhang, Liu, Tang, and Coates}]{wu2022adapting}
\bibinfo{author}{H.~Wu}, \bibinfo{author}{C.~Ma}, \bibinfo{author}{Y.~Zhang}, \bibinfo{author}{X.~Liu}, \bibinfo{author}{R.~Tang}, \bibinfo{author}{M.~Coates},
\newblock \bibinfo{title}{Adapting triplet importance of implicit feedback for personalized recommendation},
\newblock in: \bibinfo{booktitle}{International Conference on Information \& Knowledge Management (CIKM)}, \bibinfo{publisher}{ACM}, \bibinfo{year}{2022}, p. \bibinfo{pages}{2148–2157}. \DOIprefix\doi{https://doi.org/10.1145/3511808.3557229}.
\bibitem[{Bai et~al.(2024)Bai, Wang, Zhang, Song, Wu, and Nie}]{BAI2024103525}
\bibinfo{author}{T.~Bai}, \bibinfo{author}{X.~Wang}, \bibinfo{author}{Z.~Zhang}, \bibinfo{author}{W.~Song}, \bibinfo{author}{B.~Wu}, \bibinfo{author}{J.-Y. Nie},
\newblock \bibinfo{title}{Gpr-opt: A practical gaussian optimization criterion for implicit recommender systems},
\newblock \bibinfo{journal}{Information Processing \& Management} \bibinfo{volume}{61} (\bibinfo{year}{2024}) \bibinfo{pages}{103525}.
\bibitem[{Roy and Hasan(2023)}]{roy2023item}
\bibinfo{author}{F.~Roy}, \bibinfo{author}{M.~Hasan},
\newblock \bibinfo{title}{An item--item collaborative filtering recommender system based on item reviews: An approach with deep learning.},
\newblock \bibinfo{journal}{Vietnam Journal of Computer Science (World Scientific)} \bibinfo{volume}{10} (\bibinfo{year}{2023}).
\bibitem[{Ricci et~al.(2021)Ricci, Rokach, and Shapira}]{ricci2021recommender}
\bibinfo{author}{F.~Ricci}, \bibinfo{author}{L.~Rokach}, \bibinfo{author}{B.~Shapira},
\newblock \bibinfo{title}{Recommender systems: Techniques, applications, and challenges},
\newblock \bibinfo{journal}{Recommender Systems Handbook}  (\bibinfo{year}{2021}) \bibinfo{pages}{1--35}.
\bibitem[{Reusens et~al.(2017)Reusens, Lemahieu, Baesens, and Sels}]{reusens2017note}
\bibinfo{author}{M.~Reusens}, \bibinfo{author}{W.~Lemahieu}, \bibinfo{author}{B.~Baesens}, \bibinfo{author}{L.~Sels},
\newblock \bibinfo{title}{A note on explicit versus implicit information for job recommendation},
\newblock \bibinfo{journal}{Decision Support Systems} \bibinfo{volume}{98} (\bibinfo{year}{2017}) \bibinfo{pages}{26--35}.
\bibitem[{Deng et~al.(2025)Deng, Zhou, Haq, Ahmad, and Tabassum}]{DENG2025DPDSA}
\bibinfo{author}{Y.~Deng}, \bibinfo{author}{W.~Zhou}, \bibinfo{author}{A.~U. Haq}, \bibinfo{author}{S.~Ahmad}, \bibinfo{author}{A.~Tabassum},
\newblock \bibinfo{title}{Differentially private recommender framework with dual semi-autoencoder},
\newblock \bibinfo{journal}{Expert Systems with Applications} \bibinfo{volume}{260} (\bibinfo{year}{2025}) \bibinfo{pages}{125447}.
\bibitem[{Ghasemi and Momtazi(2021)}]{ghasemi2021neural}
\bibinfo{author}{N.~Ghasemi}, \bibinfo{author}{S.~Momtazi},
\newblock \bibinfo{title}{Neural text similarity of user reviews for improving collaborative filtering recommender systems},
\newblock \bibinfo{journal}{Electronic Commerce Research and Applications} \bibinfo{volume}{45} (\bibinfo{year}{2021}) \bibinfo{pages}{101019}.
\bibitem[{Wang et~al.(2021)Wang, Zuo, Li, and Wu}]{wang2021cross}
\bibinfo{author}{H.~Wang}, \bibinfo{author}{Y.~Zuo}, \bibinfo{author}{H.~Li}, \bibinfo{author}{J.~Wu},
\newblock \bibinfo{title}{Cross-domain recommendation with user personality},
\newblock \bibinfo{journal}{Knowledge-Based Systems} \bibinfo{volume}{213} (\bibinfo{year}{2021}) \bibinfo{pages}{106664}.
\bibitem[{Cao et~al.(2022)Cao, Cong, Sheng, Liu, and Wang}]{cao2022contrastive}
\bibinfo{author}{J.~Cao}, \bibinfo{author}{X.~Cong}, \bibinfo{author}{J.~Sheng}, \bibinfo{author}{T.~Liu}, \bibinfo{author}{B.~Wang},
\newblock \bibinfo{title}{Contrastive cross-domain sequential recommendation},
\newblock in: \bibinfo{booktitle}{International Conference on Information \& Knowledge Management (CIKM)}, \bibinfo{publisher}{ACM}, \bibinfo{year}{2022}, pp. \bibinfo{pages}{138--147}. \DOIprefix\doi{https://doi.org/10.1145/3511808.3557262}.
\bibitem[{Khan et~al.(2020)Khan, Mukta, Ali, and Mahmud}]{khan2020predicting}
\bibinfo{author}{E.~M. Khan}, \bibinfo{author}{M.~S.~H. Mukta}, \bibinfo{author}{M.~E. Ali}, \bibinfo{author}{J.~Mahmud},
\newblock \bibinfo{title}{Predicting users’ movie preference and rating behavior from personality and values},
\newblock \bibinfo{journal}{ACM Transactions on Interactive Intelligent Systems (TiiS)} \bibinfo{volume}{10} (\bibinfo{year}{2020}) \bibinfo{pages}{1--25}.
\bibitem[{Slokom et~al.(2021)Slokom, Hanjalic, and Larson}]{base01_2021Gender}
\bibinfo{author}{M.~Slokom}, \bibinfo{author}{A.~Hanjalic}, \bibinfo{author}{M.~Larson},
\newblock \bibinfo{title}{Towards user-oriented privacy for recommender system data: A personalization-based approach to gender obfuscation for user profiles},
\newblock \bibinfo{journal}{Information Processing \& Management} \bibinfo{volume}{58} (\bibinfo{year}{2021}) \bibinfo{pages}{102722}.
\bibitem[{Weinsberg et~al.(2012)Weinsberg, Bhagat, Ioannidis, and Taft}]{weinsberg2012blurme}
\bibinfo{author}{U.~Weinsberg}, \bibinfo{author}{S.~Bhagat}, \bibinfo{author}{S.~Ioannidis}, \bibinfo{author}{N.~Taft},
\newblock \bibinfo{title}{Blurme: Inferring and obfuscating user gender based on ratings},
\newblock in: \bibinfo{booktitle}{ACM conference on Recommender Systems (RecSys)}, \bibinfo{publisher}{ACM}, \bibinfo{year}{2012}, pp. \bibinfo{pages}{195--202}. \DOIprefix\doi{https://doi.org/10.1145/2365952.2365989}.
\bibitem[{W{\"u}hr et~al.(2017)W{\"u}hr, Lange, and Schwarz}]{wuhr2017tears}
\bibinfo{author}{P.~W{\"u}hr}, \bibinfo{author}{B.~P. Lange}, \bibinfo{author}{S.~Schwarz},
\newblock \bibinfo{title}{Tears or fears? comparing gender stereotypes about movie preferences to actual preferences},
\newblock \bibinfo{journal}{Frontiers in psychology} \bibinfo{volume}{8} (\bibinfo{year}{2017}) \bibinfo{pages}{249605}.
\bibitem[{Russo and Stol(2020)}]{russo2020gender}
\bibinfo{author}{D.~Russo}, \bibinfo{author}{K.-J. Stol},
\newblock \bibinfo{title}{Gender differences in personality traits of software engineers},
\newblock \bibinfo{journal}{IEEE Transactions on Software Engineering} \bibinfo{volume}{48} (\bibinfo{year}{2020}) \bibinfo{pages}{819--834}.
\bibitem[{Fabris et~al.(2020)Fabris, Purpura, Silvello, and Susto}]{fabris2020gender}
\bibinfo{author}{A.~Fabris}, \bibinfo{author}{A.~Purpura}, \bibinfo{author}{G.~Silvello}, \bibinfo{author}{G.~A. Susto},
\newblock \bibinfo{title}{Gender stereotype reinforcement: Measuring the gender bias conveyed by ranking algorithms},
\newblock \bibinfo{journal}{Information Processing \& Management} \bibinfo{volume}{57} (\bibinfo{year}{2020}) \bibinfo{pages}{102377}.
\bibitem[{Ashton-James et~al.(2019)Ashton-James, Tybur, Grie{\ss}er, and Costa}]{ashton2019stereotypes}
\bibinfo{author}{C.~E. Ashton-James}, \bibinfo{author}{J.~M. Tybur}, \bibinfo{author}{V.~Grie{\ss}er}, \bibinfo{author}{D.~Costa},
\newblock \bibinfo{title}{Stereotypes about surgeon warmth and competence: the role of surgeon gender},
\newblock \bibinfo{journal}{PLoS One} \bibinfo{volume}{14} (\bibinfo{year}{2019}) \bibinfo{pages}{e0211890}.
\bibitem[{Eyssel and Hegel(2012)}]{eyssel2012s}
\bibinfo{author}{F.~Eyssel}, \bibinfo{author}{F.~Hegel},
\newblock \bibinfo{title}{(s) he's got the look: Gender stereotyping of robots 1},
\newblock \bibinfo{journal}{Journal of Applied Social Psychology} \bibinfo{volume}{42} (\bibinfo{year}{2012}) \bibinfo{pages}{2213--2230}.
\bibitem[{Wynn and Correll(2018)}]{wynn2018combating}
\bibinfo{author}{A.~T. Wynn}, \bibinfo{author}{S.~J. Correll},
\newblock \bibinfo{title}{Combating gender bias in modern workplaces},
\newblock \bibinfo{journal}{Handbook of the Sociology of Gender}  (\bibinfo{year}{2018}) \bibinfo{pages}{509--521}.
\bibitem[{Ahn et~al.(2022)Ahn, Kim, and Sung}]{ahn2022effect}
\bibinfo{author}{J.~Ahn}, \bibinfo{author}{J.~Kim}, \bibinfo{author}{Y.~Sung},
\newblock \bibinfo{title}{The effect of gender stereotypes on artificial intelligence recommendations},
\newblock \bibinfo{journal}{Journal of Business Research} \bibinfo{volume}{141} (\bibinfo{year}{2022}) \bibinfo{pages}{50--59}.
\bibitem[{Lange et~al.(2021)Lange, W{\"u}hr, and Schwarz}]{lange2021time}
\bibinfo{author}{B.~P. Lange}, \bibinfo{author}{P.~W{\"u}hr}, \bibinfo{author}{S.~Schwarz},
\newblock \bibinfo{title}{Of time gals and mega men: Empirical findings on gender differences in digital game genre preferences and the accuracy of respective gender stereotypes},
\newblock \bibinfo{journal}{Frontiers in Psychology} \bibinfo{volume}{12} (\bibinfo{year}{2021}) \bibinfo{pages}{657430}.
\bibitem[{Salamatian et~al.(2015)Salamatian, Zhang, du~Pin~Calmon, Bhamidipati, Fawaz, Kveton, Oliveira, and Taft}]{salamatian2015managing}
\bibinfo{author}{S.~Salamatian}, \bibinfo{author}{A.~Zhang}, \bibinfo{author}{F.~du~Pin~Calmon}, \bibinfo{author}{S.~Bhamidipati}, \bibinfo{author}{N.~Fawaz}, \bibinfo{author}{B.~Kveton}, \bibinfo{author}{P.~Oliveira}, \bibinfo{author}{N.~Taft},
\newblock \bibinfo{title}{Managing your private and public data: Bringing down inference attacks against your privacy},
\newblock \bibinfo{journal}{IEEE Journal of Selected Topics in Signal Processing} \bibinfo{volume}{9} (\bibinfo{year}{2015}) \bibinfo{pages}{1240--1255}.
\bibitem[{Kosinski et~al.(2013)Kosinski, Stillwell, and Graepel}]{kosinski2013private}
\bibinfo{author}{M.~Kosinski}, \bibinfo{author}{D.~Stillwell}, \bibinfo{author}{T.~Graepel},
\newblock \bibinfo{title}{Private traits and attributes are predictable from digital records of human behavior},
\newblock \bibinfo{journal}{Proceedings of the national academy of sciences} \bibinfo{volume}{110} (\bibinfo{year}{2013}) \bibinfo{pages}{5802--5805}.
\bibitem[{Bi et~al.(2013)Bi, Shokouhi, Kosinski, and Graepel}]{bi2013inferring}
\bibinfo{author}{B.~Bi}, \bibinfo{author}{M.~Shokouhi}, \bibinfo{author}{M.~Kosinski}, \bibinfo{author}{T.~Graepel},
\newblock \bibinfo{title}{Inferring the demographics of search users: Social data meets search queries},
\newblock in: \bibinfo{booktitle}{International Conference on World Wide Web (WWW)}, \bibinfo{publisher}{ACM}, \bibinfo{year}{2013}, pp. \bibinfo{pages}{131--140}. \DOIprefix\doi{https://doi.org/10.1145/2488388.2488401}.
\bibitem[{Feng et~al.(2014)Feng, Guo, Chen, Tan, Xu, Shen, and Zhu}]{feng2014tags}
\bibinfo{author}{T.~Feng}, \bibinfo{author}{Y.~Guo}, \bibinfo{author}{Y.~Chen}, \bibinfo{author}{X.~Tan}, \bibinfo{author}{T.~Xu}, \bibinfo{author}{B.~Shen}, \bibinfo{author}{W.~Zhu},
\newblock \bibinfo{title}{Tags and titles of videos you watched tell your gender},
\newblock in: \bibinfo{booktitle}{IEEE International Conference on Communications (ICC)}, \bibinfo{organization}{IEEE}, \bibinfo{year}{2014}, pp. \bibinfo{pages}{1837--1842}. \DOIprefix\doi{https://doi.org/10.1109/ICC.2014.6883590}.
\bibitem[{Jia et~al.(2017)Jia, Wang, Zhang, and Gong}]{jia2017attriinfer}
\bibinfo{author}{J.~Jia}, \bibinfo{author}{B.~Wang}, \bibinfo{author}{L.~Zhang}, \bibinfo{author}{N.~Z. Gong},
\newblock \bibinfo{title}{Attriinfer: Inferring user attributes in online social networks using markov random fields},
\newblock in: \bibinfo{booktitle}{International Conference on World Wide Web (WWW)}, \bibinfo{publisher}{International World Wide Web Conferences Steering Committee}, \bibinfo{year}{2017}, pp. \bibinfo{pages}{1561--1569}. \DOIprefix\doi{https://doi.org/10.1145/3038912.3052695}.
\bibitem[{Feng et~al.(2015)Feng, Guo, and Chen}]{feng2015can}
\bibinfo{author}{T.~Feng}, \bibinfo{author}{Y.~Guo}, \bibinfo{author}{Y.~Chen},
\newblock \bibinfo{title}{Can user privacy and recommendation performance be preserved simultaneously?},
\newblock \bibinfo{journal}{Computer Communications} \bibinfo{volume}{68} (\bibinfo{year}{2015}) \bibinfo{pages}{17--24}.
\bibitem[{Correia and Barbosa(2018)}]{correia2018cinema}
\bibinfo{author}{A.~F. Correia}, \bibinfo{author}{S.~Barbosa},
\newblock \bibinfo{title}{Cinema, aesthetics and narrative: Cinema as therapy in substance use disorders},
\newblock \bibinfo{journal}{The Arts in Psychotherapy} \bibinfo{volume}{60} (\bibinfo{year}{2018}) \bibinfo{pages}{63--71}.
\bibitem[{Palomba(2020)}]{palomba2020consumer}
\bibinfo{author}{A.~Palomba},
\newblock \bibinfo{title}{Consumer personality and lifestyles at the box office and beyond: How demographics, lifestyles and personalities predict movie consumption},
\newblock \bibinfo{journal}{Journal of Retailing and Consumer Services} \bibinfo{volume}{55} (\bibinfo{year}{2020}) \bibinfo{pages}{102083}.
\bibitem[{Kim(2018)}]{kim2018demographic}
\bibinfo{author}{D.~D.~E. Kim},
\newblock \bibinfo{title}{Demographic differences in perceptions of media brand personality: a multilevel analysis},
\newblock \bibinfo{journal}{International Journal on Media Management} \bibinfo{volume}{20} (\bibinfo{year}{2018}) \bibinfo{pages}{81--106}.
\bibitem[{Infortuna et~al.(2021)Infortuna, Battaglia, Freedberg, Mento, Zoccali, Muscatello, and Bruno}]{infortuna2021inner}
\bibinfo{author}{C.~Infortuna}, \bibinfo{author}{F.~Battaglia}, \bibinfo{author}{D.~Freedberg}, \bibinfo{author}{C.~Mento}, \bibinfo{author}{R.~A. Zoccali}, \bibinfo{author}{M.~R.~A. Muscatello}, \bibinfo{author}{A.~Bruno},
\newblock \bibinfo{title}{The inner muses: How affective temperament traits, gender and age predict film genre preference},
\newblock \bibinfo{journal}{Personality and Individual Differences} \bibinfo{volume}{178} (\bibinfo{year}{2021}) \bibinfo{pages}{110877}.
\bibitem[{Martin(2019)}]{martin2019you}
\bibinfo{author}{G.~N. Martin},
\newblock \bibinfo{title}{(why) do you like scary movies? a review of the empirical research on psychological responses to horror films},
\newblock \bibinfo{journal}{Frontiers in psychology} \bibinfo{volume}{10} (\bibinfo{year}{2019}) \bibinfo{pages}{430538}.
\bibitem[{Grodal(2017)}]{grodal2017film}
\bibinfo{author}{T.~Grodal},
\newblock \bibinfo{title}{How film genres are a product of biology, evolution and culture—an embodied approach},
\newblock \bibinfo{journal}{Palgrave Communications} \bibinfo{volume}{3} (\bibinfo{year}{2017}) \bibinfo{pages}{1--8}.
\bibitem[{Procyshyn et~al.(2020)Procyshyn, Watson, and Crespi}]{procyshyn2020experimental}
\bibinfo{author}{T.~L. Procyshyn}, \bibinfo{author}{N.~V. Watson}, \bibinfo{author}{B.~J. Crespi},
\newblock \bibinfo{title}{Experimental empathy induction promotes oxytocin increases and testosterone decreases},
\newblock \bibinfo{journal}{Hormones and behavior} \bibinfo{volume}{117} (\bibinfo{year}{2020}) \bibinfo{pages}{104607}.
\bibitem[{Steck(2018)}]{steck2018calibrated}
\bibinfo{author}{H.~Steck},
\newblock \bibinfo{title}{Calibrated recommendations},
\newblock in: \bibinfo{booktitle}{ACM conference on Recommender Systems (RecSys)}, \bibinfo{publisher}{ACM}, \bibinfo{year}{2018}, pp. \bibinfo{pages}{154--162}. \DOIprefix\doi{https://doi.org/10.1145/3240323.3240372}.
\bibitem[{Bolock et~al.(2020)Bolock, Kady, Herbert, and Abdennadher}]{bolock2020towards}
\bibinfo{author}{A.~E. Bolock}, \bibinfo{author}{A.~E. Kady}, \bibinfo{author}{C.~Herbert}, \bibinfo{author}{S.~Abdennadher},
\newblock \bibinfo{title}{Towards a character-based meta recommender for movies},
\newblock in: \bibinfo{booktitle}{Computational Science and Technology: 6th ICCST 2019, Kota Kinabalu, Malaysia, 29-30 August 2019}, \bibinfo{organization}{Springer}, \bibinfo{year}{2020}, pp. \bibinfo{pages}{627--638}. \DOIprefix\doi{https://doi.org/10.1007/978-981-15-0058-9_60}.
\bibitem[{Afsar et~al.(2022)Afsar, Crump, and Far}]{afsar2022reinforcement}
\bibinfo{author}{M.~M. Afsar}, \bibinfo{author}{T.~Crump}, \bibinfo{author}{B.~Far},
\newblock \bibinfo{title}{Reinforcement learning based recommender systems: A survey},
\newblock \bibinfo{journal}{ACM Computing Surveys} \bibinfo{volume}{55} (\bibinfo{year}{2022}) \bibinfo{pages}{1--38}.
\bibitem[{Hasan and Roy(2019)}]{hasan2019item}
\bibinfo{author}{M.~Hasan}, \bibinfo{author}{F.~Roy},
\newblock \bibinfo{title}{An item-item collaborative filtering recommender system using trust and genre to address the cold-start problem},
\newblock \bibinfo{journal}{Big Data and Cognitive Computing} \bibinfo{volume}{3} (\bibinfo{year}{2019}) \bibinfo{pages}{39}.
\bibitem[{Zhu et~al.(2017)Zhu, Baesens, and vanden Broucke}]{zhu2017empirical}
\bibinfo{author}{B.~Zhu}, \bibinfo{author}{B.~Baesens}, \bibinfo{author}{S.~K. vanden Broucke},
\newblock \bibinfo{title}{An empirical comparison of techniques for the class imbalance problem in churn prediction},
\newblock \bibinfo{journal}{Information sciences} \bibinfo{volume}{408} (\bibinfo{year}{2017}) \bibinfo{pages}{84--99}.
\bibitem[{Ekstrand et~al.(2018)Ekstrand, Tian, Kazi, Mehrpouyan, and Kluver}]{ekstrand2018exploring}
\bibinfo{author}{M.~D. Ekstrand}, \bibinfo{author}{M.~Tian}, \bibinfo{author}{M.~R.~I. Kazi}, \bibinfo{author}{H.~Mehrpouyan}, \bibinfo{author}{D.~Kluver},
\newblock \bibinfo{title}{Exploring author gender in book rating and recommendation},
\newblock in: \bibinfo{booktitle}{ACM Conference on Recommender Systems (RecSys)}, \bibinfo{publisher}{ACM}, \bibinfo{year}{2018}, pp. \bibinfo{pages}{242--250}. \DOIprefix\doi{https://doi.org/10.1145/3240323.3240373}.
\bibitem[{Melchiorre et~al.(2021)Melchiorre, Rekabsaz, Parada-Cabaleiro, Brandl, Lesota, and Schedl}]{melchiorre2021investigating}
\bibinfo{author}{A.~B. Melchiorre}, \bibinfo{author}{N.~Rekabsaz}, \bibinfo{author}{E.~Parada-Cabaleiro}, \bibinfo{author}{S.~Brandl}, \bibinfo{author}{O.~Lesota}, \bibinfo{author}{M.~Schedl},
\newblock \bibinfo{title}{Investigating gender fairness of recommendation algorithms in the music domain},
\newblock \bibinfo{journal}{Information Processing \& Management} \bibinfo{volume}{58} (\bibinfo{year}{2021}) \bibinfo{pages}{102666}.
\bibitem[{G{\'o}mez et~al.(2022)G{\'o}mez, Boratto, and Salam{\'o}}]{gomez2022provider}
\bibinfo{author}{E.~G{\'o}mez}, \bibinfo{author}{L.~Boratto}, \bibinfo{author}{M.~Salam{\'o}},
\newblock \bibinfo{title}{Provider fairness across continents in collaborative recommender systems},
\newblock \bibinfo{journal}{Information Processing \& Management} \bibinfo{volume}{59} (\bibinfo{year}{2022}) \bibinfo{pages}{102719}.
\bibitem[{Mansoury et~al.(2020)Mansoury, Abdollahpouri, Pechenizkiy, Mobasher, and Burke}]{mansoury2020feedback}
\bibinfo{author}{M.~Mansoury}, \bibinfo{author}{H.~Abdollahpouri}, \bibinfo{author}{M.~Pechenizkiy}, \bibinfo{author}{B.~Mobasher}, \bibinfo{author}{R.~Burke},
\newblock \bibinfo{title}{Feedback loop and bias amplification in recommender systems},
\newblock in: \bibinfo{booktitle}{ACM International Conference on Information \& Knowledge Management (CIKM)}, \bibinfo{publisher}{ACM}, \bibinfo{year}{2020}, pp. \bibinfo{pages}{2145--2148}. \DOIprefix\doi{https://doi.org/10.1145/3340531.3412152}.

\end{thebibliography}

\end{document}